\pdfoutput=1

\documentclass[12pt,preprint]{aastex}

\newcommand{\mdot}{${\dot{\rm M}}$}
\newcommand{\msun}{${{\rm M}_\odot}$}

\newcommand{\etal}{{et~al.}~}

\def\arcsec{\hbox{$^{\prime\prime}$~}}

\shorttitle{Star Formation in BCGs with GALEX}
\shortauthors{Hicks, Mushotzky \& Donahue}

\begin{document}

\title{Detecting Star Formation in Brightest Cluster Galaxies with GALEX\footnote{Based on observations made with the NASA Galaxy Evolution Explorer. GALEX is operated for NASA by the California Institute of Technology under NASA contract NAS5-98034.}}
\author{A. K. Hicks\altaffilmark{1}}
\email{ahicks@alum.mit.edu}

\author{R. Mushotzky\altaffilmark{2,3}} 
\email{richard@astro.umd.edu}
\and
\author{M. Donahue\altaffilmark{1}} 
\email{donahue@pa.msu.edu}

\altaffiltext{1}{Department of Physics \& Astronomy, Michigan State University, East Lansing, MI 48824-2320}
\altaffiltext{2}{Goddard Space Flight Center, Code 662, Greenbelt, MD, 20771} 
\altaffiltext{3}{Department of Astronomy, University of Maryland, College Park, MD 20742-2421 }

\begin{abstract}
We present the results of GALEX observations of 17 cool core (CC) clusters of galaxies. We show that GALEX is easily capable of detecting star formation in brightest cluster galaxies (BCGs) out to $z\ge 0.45$ and 50-100 kpc. In most of the CC clusters studied, we find significant UV luminosity excesses and colors that strongly suggest recent and/or current star formation. The BCGs are found to have blue UV colors in the center that become increasingly redder with radius, indicating that the UV signature of star formation is most easily detected in the central regions. Our findings show good agreement between UV star formation rates and estimates based on H$\alpha$ observations. IR observations coupled with our data indicate moderate-to-high dust attenuation. Comparisons between our UV results and the X-ray properties of our sample suggest clear correlations between UV excess, cluster entropy, and central cooling time, confirming that the star formation is directly and incontrovertibly related to the cooling gas.
\end{abstract}


\section{Introduction\label{s:intro}}

The existence of star formation in ``cooling flow'' (hereafter referred to as cool core, or CC) clusters has been a contentious issue for over 25 years \citep{fabian86}. A class of clusters with high central gas densities and theoretically short cooling times was discovered with the Einstein observatory~\citep[e.g.,][]{fabian77,cowie77,mathews78,canizares79, mushotzky81} 
via X-ray imaging and low resolution X-ray spectroscopy. These objects were also associated with H$\alpha$ filaments \citep[e.g.,][]{cowie83,heckman89}, central radio sources, high Faraday rotation \citep[e.g.,][]{ge94}, extra blue light  \citep[e.g.,][]{mcnamara89,cardiel98} and spatial coincidence of the X-ray peak with the central radio source~\citep[e.g.,][]{burns81}.  The simplest physical model was one in which the gas in the center cooled by radiating away its thermal energy, gradually losing pressure support, resulting in a flow~\citep[e.g.,][]{cowie77,fabian77,mathews78}. Cool cores are inferred to be present in more than half of all clusters at low-redshift~\citep{peres98}, and nearly as prevalent
at moderate-z~\citep[$30-50\%$ at $0.15 < z < 0.4$;][]{bauer05}.

It has become clear with the advent of XMM and Chandra data that almost every such cluster also shows a temperature drop in the center \citep[e.g.,][]{cavagnolo08}. However, measurements with the high spectral resolution XMM RGS spectrometer \citep[e.g.,][]{peterson03, kaastra04, piffaretti06} show that the X-ray spectra of the cooler gas has major differences from the theoretical cooling flow model, with a marked absence of gas at temperatures below $\sim1/3$ of the average cluster temperature. Thus it remains a mystery what happens to the cool gas. The combination of Chandra imaging and radio data \citep{mcnamara00,blanton01} have shown that in many of these objects there exist holes in the X-ray surface brightness which are filled in by radio emitting plasma, thus lending credence to ideas that feedback from AGN in brightest cluster galaxies strongly modifies the cooling and thus reduces the net amount of material available for star formation.  

Despite past evidence of star formation in these systems \citep[e.g.,][]{mcnamara89,cardiel98,crawford99} and the apparent preference for emission-line systems to inhabit the cores of high-central density, short cooling time clusters~\citep{hu85}, the fact that star formation rate (SFR) estimates often differed by factors of $\sim10$ or more from inferred X-ray cooling rates led to doubt that the two phenomena were related. However, recent UV investigations \citep{mittaz01,hicks05}, Spitzer data \citep{quillen08,odea08}, and precision optical photometry \citep{bildfell08} have definitively shown that CC clusters are indeed the sites of star formation, and that there is an indisputable relationship between X-ray properties and SFRs. 

Previous studies have shown a connection between activity, such as star formation and radio AGN, in brightest cluster galaxies (BCGs) and the thermodynamic state (traced by entropy, cooling time, etc.) of the intracluster medium in the cluster core~\citep[e.g.,][]{cavagnolo08}. The physical explanation for a connection between the state of the hot gas in the core inside 100 kpc and star formation in the inner 10 kpc in the brightest cluster galaxy is not at all obvious. However the situation in nearby BCGs might be similar to what models predict is happening in massive galaxies at high redshift. Models such as \citet{ostricker05} suggest that almost all star formation in the high-redshift universe derives from accretion of gas that fell into potential wells, shocked, heated, then cooled~\citep{white91, fabian86}. More recent simulations~\citep[e.g.,][]{keres09} show that this ``hot mode'' of accretion might be the dominant mode of star formation for massive galaxies, but this verdict is far from final owing to the unknown effects of feedback. 

In either steady state or bursty star formation scenarios the  dominant contribution to the UV flux comes from short-lived main sequence stars, therefore the UV band constitutes one of the prime routes to understanding star formation. Despite this fact, while cool core clusters have been well observed in the radio \citep[e.g.,][]{mcnamara00,blanton01}, IR \citep[e.g.,][]{quillen08,odea08} and with emission line studies \citep[e.g.,][]{heckman89}, there have been comparatively few published studies of UV observations of cool core clusters~\citep{odea04,hicks05,sparks09,donahue10}. Here we attempt to address and quantify the connection between X-ray properties and star formation, using recent GALEX observations of a sample of 16 CC clusters. 
Unless otherwise noted, this paper assumes a cosmology of $\rm{H}_0=70~\rm{km}~\rm{s}^{-1}~\rm{Mpc}^{-1}$, $\Omega_{\rm{M}}=0.3$, and $\Omega_{\Lambda}=0.7$

\section{Observations \label{data}}

The Galaxy Evolution Explorer (GALEX) is an orbiting space telescope possessing both imaging and spectroscopic capabilities in two ultraviolet wavebands, Far UV (FUV) 1350-1780~\AA ~and Near UV (NUV) 1770-2730~\AA ~\citep{martin05}. GALEX has a very low sky background, and very high sensitivity ($\sim24$ apparent AB magnitudes in each filter for 2 ksec exposures). Pixels are $1.5\arcsec\times~1.5\arcsec$, and GALEX's on-axis spatial resolution is 4.2$\arcsec$ and 4.9$\arcsec$ for the FUV and NUV respectively (GALEX Technical Documentation\footnote{
\url{http://www.galex.caltech.edu/researcher/techdocs.html}}).

\begin{figure}
\centerline{\includegraphics[width=6.in]{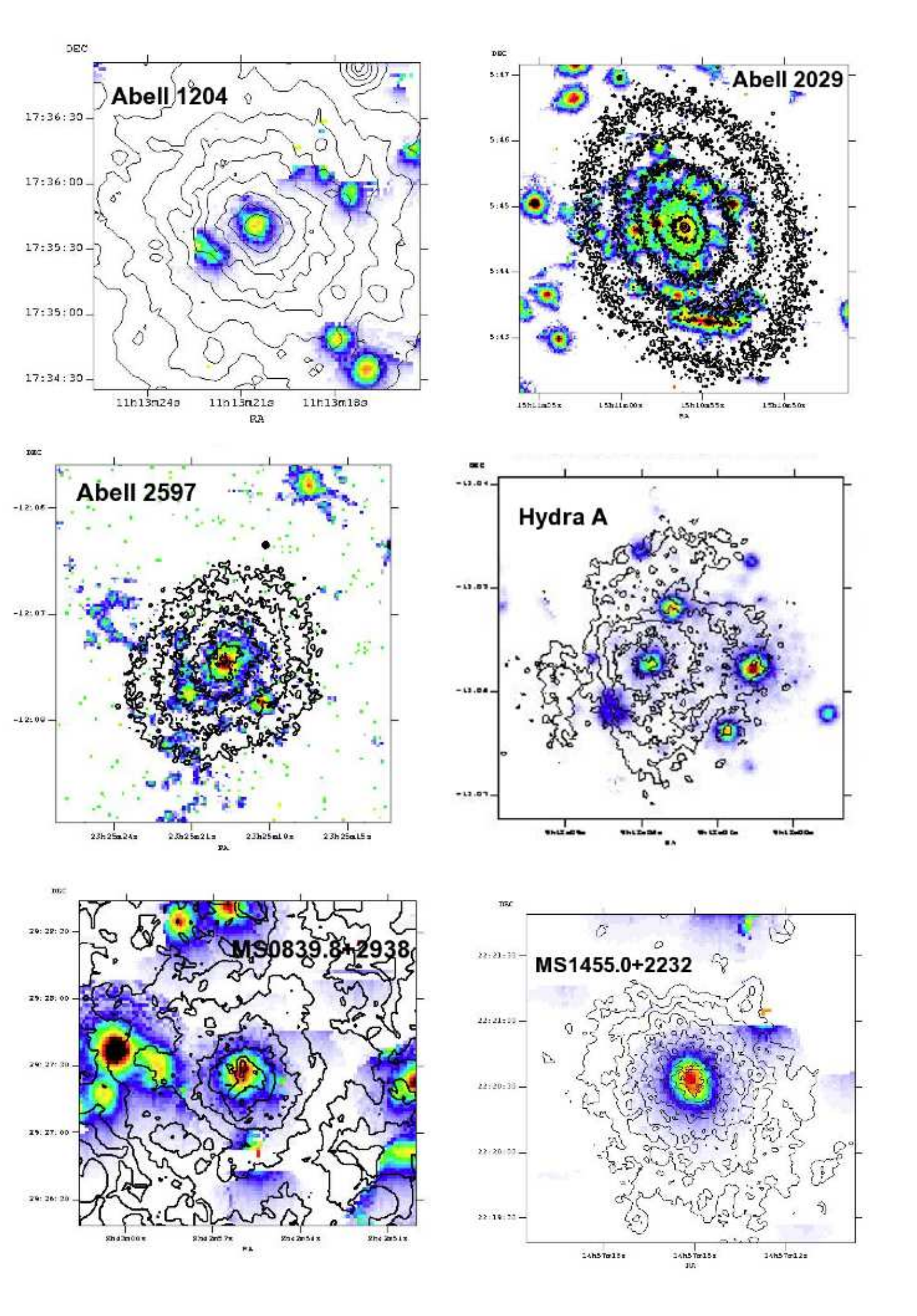}}
\caption{GALEX NUV images with X-ray contours overlaid, shown for a representative subset of our sample. Note that the UV emission is generally well aligned with the densest X-ray emitting gas.\label{fig1}}
\end{figure}

Our GALEX targets consist of 17 clusters of galaxies that exhibit evidence of central cooling based on the indicators discussed in the introduction.  These objects were chosen to sample a wide range of redshifts ($0.02<z<0.45$) and central cooling times ($0.5 < \rm{t}_{\rm{cool}} < 4.6$ Gyr at R $=20$ kpc), be safely observable by GALEX (i.e., no bright nearby UV sources) and have low attenuations (Av$ < 0.5$); therefore they do not constitute a complete sample. One of our targets (Abell 644) was observed after the loss of the GALEX FUV detector and therefore only has NUV data. Figure \ref{fig1} shows the GALEX NUV images of a representative sample of our targets with {\it{Chandra}} X-ray contours overlayed.  Table \ref{table1} lists the objects in our sample, their redshifts, and GALEX exposure times. These exposure times were based on our previous work with XMM-Newton Optical Monitor UV data~\citep{hicks05}.

\section{Photometry\label{phot}}

All photometry was performed on pipeline-processed GALEX intensity maps. FUV data were convolved with a gaussian to match the NUV PSF, and the sizes of point sources in the resulting images were checked against those in the NUV data with the IRAF tool {\it{imexam}}. Photon counts were corrected for background using large ($\sim40$\arcsec radius) source-free regions taken from nearby areas of the camera in the same observation. Our photometric measurements were compared to those obtained with the background-subtracted intensity images provided in the GALEX pipeline, as an added check.

Final fluxes were determined by employing GALEX counts-to-flux conversions, and correcting for average Galactic extinction in the line of sight to each cluster \citep{cardelli}. Errors were assessed by adding Poisson (root N) photon statistics in quadrature to a conservative 5\% fixed systematic error~\citep{morrissey07}. 
All of our targets were easily detected in both GALEX wavebands, with an average SNR of 40 (21) in the NUV (FUV), and minimum SNRs of $\sim6$ in each band (for a fixed 7\arcsec radius aperture). Photometric results are presented in Table~\ref{table2}.

\section{Spatial Analysis\label{profiles}}

To investigate the spatial distribution of UV emission in our targets, radial flux profiles were produced for each band from point source subtracted intensity maps (convolved to produce matching PSFs, as above). Profiles were constructed using concentric annuli of at least 5\arcsec width and binned to achieve S/N $> 3$.  These profiles are shown in Figure~\ref{fig2}. With GALEX we are able to detect UV emission out to large  radii in many of the CC clusters. 

The surface brightness profiles were then background subtracted and combined to create radial color profiles for each cluster (Figure~\ref{fig3}). Overall the central colors of most of the BCGs
indicate the presence of a very young stellar population, and imply active star-formation. We also see positive color gradients in nearly all of our targets, in keeping with the results of ~\citet{rafferty08} and~\citet{wang10} for cool core clusters.  

Greater than $82\%$ of elliptical galaxies have FUV-NUV colors of $>0.9$~\citep[][]{gildepaz07}, much redder than the central regions of all of our objects (Figure~\ref{fig3}). However, presumably due to variations in the UV upturn (thought to be caused by horizontal branch stars) there is still a broad distribution of UV color among passive ellipticals, therefore we have not attempted to determine a definitive extent of star formation in individual targets.

\section{Fixed Aperture Analysis}

To determine the amount of ``excess'' UV light present in our targets, we first need an estimate of their ``expected'' UV emission. We obtain this empirically by examining the UV emission of non-star forming ellipticals and BCGs, using 2MASS J band flux as a proxy for the old stellar population. We avail ourselves of existing 2MASS photometric measurements by adopting their fixed 7\arcsec radius aperture in this portion of our analysis. We note that this aperture contains the majority of excess UV emission (Figures~\ref{fig2} and~\ref{fig3}).

\subsection{Control Sample\label{cal}}

Our non-star forming control sample is composed of 17 cluster ellipticals and 22 BCGs in non-CC clusters, all drawn from archival GALEX observations. The clusters used in our calibration analysis are listed in Table~\ref{table3} along with their redshifts and GALEX exposure times.  

Elliptical galaxies were gathered from 4 clusters spanning a redshift range of $0.08 < z < 0.15$. We used FUV-K colors as a proxy for galaxy type~\citep{gildepaz07}, adopting a liberal cutoff of FUV-K~$= 7.5$.

BCGs were selected for inclusion if they met one of the following criteria: 1) central cooling time $>7$ Gyr; 2) spectrally determined X-ray \mdot~consistent with zero; 3) X-ray underluminous~\citep{popesso07} and 1.4 GHz luminosity $<1 \times 10^{24}~\rm{W}~\rm{Hz}^{-1}$~\citep[][]{sun09}. BCGs without ancillary X-ray or radio data were included when FUV-K~$ > 7.5$.  

Our total (elliptical + BCG) calibration sample spans ranges in redshift and absolute J magnitude that are well-matched to our target sample, with the exceptions of our highest-redshift cluster (RXJ1347.5-1145 at z=0.45) and most luminous BCG (MKW4).  We note that none of our conclusions are based on individual objects in our sample.

Photometry was performed in $7\arcsec$ radius apertures centered on each galaxy (as described in Section~\ref{phot}); measurements are given in Table~\ref{table4}. We see no relationship between NUV-J color and redshift in our calibration sample, confirming that our aperture choice is sufficient for meeting the goals of this study; e.g., our aperture choice is large enough that we capture most of the ``excess'' UV light even at low redshifts and is not so large as to dilute the signal below detection thresholds at high redshifts.

Least squares fits were executed between the properties of our calibration sample using the wls\_regress algorithm of~\citet{akritas96}. This routine was chosen because the scatter in UV luminosity (ostensibly stemming from variations in the UV upturn) vastly dominates over J band luminosity uncertainties.  Relationships are fitted with the form

\begin{equation}
\label{eq:powerlaw}
        {\rm{log_{10}}}~Y = C_1 + C_2~ {\rm{log_{10}}}~X
\end{equation}


Correlations between luminosities and flux ratios are given in Table~\ref{table5} and are shown in Figures~\ref{fig4}~and~\ref{fig5}. Clearly the FUV shows more scatter than the NUV. This scatter is expected because the FUV filter covers a spectral region which is very sensitive to variations in the magnitude of the UV upturn from object to object. This high sensitivity to the UV upturn makes the NUV filter a more straightforward choice for investigating star formation in early-type galaxies, so our discussion will focus on the NUV results.

\subsection{UV Excesses}

The UV luminosity excesses of our 17 cool core clusters were calculated by subtracting the expected UV luminosity from the old stellar population (based on the fit obtained in Section \ref{cal}) from the measured value. The majority of our sample exhibits clear UV excesses, indicating recent star formation. Figure~\ref{fig5} shows the J band luminosity of each galaxy in our sample plotted against its UV/J flux ratios (UV-J colors). We have not attempted to estimate and correct for internal dust absorption, and thus these excesses provide a lower limit on the UV emission in these clusters. Our measured UV excesses are given in Table~\ref{table6}. 



Starburst99 \citep{leitherer} redshifted models corresponding to continuous Saltpeter star formation over a 20 Myr period were used to estimate star formation rates for our sample. The 20 Myr continuous model was chosen to grossly approximate episodic cooling timescales, during which the system undergoes feedback processes with alternating heating and cooling cycles; it is this model that we use in the figures to follow. We emphasize that there are too many unknowns (e.g., internal reddening, IMF, continuous vs. burst star formation, age of star formation, etc.) to predict accurate star formation rates, and that we estimate SFRs with this model purely to facilitate comparisons with previous work and with other wavebands. Resulting values are shown in Table~\ref{table6}.

Star formation rates estimated from the NUV and FUV bands show general agreement (Figure~\ref{fig6}), though FUV derived SFRs tend to be slightly lower, ostensibly due to the larger scatter in the FUV calibration relationship. Because of the overall agreement between SFR estimates in the two bands, and the tendency for NUV data to be less plagued by variations in the UV upturn, we focus primarily on the NUV results in the following sections.

\subsection{UV Colors}

Some of our targets have very blue FUV-NUV colors when compared to our control sample, but none of them are known to harbor a central AGN. It is possible that their dust is of the Milky Way variety, which preferentially absorbs NUV emission and therefore results in bluer observed colors ~\citep[e.g.,][]{witt00}.

In Figure~\ref{fig7}, we show that the FUV-NUV colors of the central (7\arcsec) of our sample are inversely correlated with excess UV luminosity (correlation coefficient $= -0.74$), such that the reddest UV colors are associated with the weakest UV excesses. Typical FUV-NUV colors for inactive BCGs are shaded grey on this plot, consistent with the colors of the BCGs with the smallest excesses.  
A couple of the BCGs' colors are undoubtedly affected by Lyman-$\alpha$ emission contributing to the FUV band (ZwCl 3146) or NUV band (RXJ1347.5-1145). 

If extinction by Milky Way-type dust is the explanation for the color trend, then the conclusion from this plot is that the BCGs with the largest UV excesses have the largest intrinsic dust extinction.  If the color differences are intrinsic to the stellar population, then the BCGs with the largest excesses have more hot main sequence stars and therefore may host more recent bursts than galaxies with lower UV excesses. 

Using our measured UV excesses, we can estimate the color of the young stellar population in our targets (ranging from -0.7 to 0.5). All but two (MKW4 and MS1358.4+6245, which are bluer than -0.3) can be explained using either continuous or burst models for the star formation. Interestingly, objects with colors redder than $\sim0.1$ (about half of the sample) can only be explained by a burst of star formation occurring 30-200 Myr ago.

\section{Multiwavelength Comparisons}

\subsection{H$\alpha$}

H$\alpha$ measurements for our sample were taken from the literature (specific references are given in Table~\ref{table7}), and are shown vs. 
NUV inferred SFR in Figure~\ref{fig8}. We note that H$\alpha$ measurements are usually based on long-slit estimates, and may miss emission line flux outside of the slit. Overlaid on the plot is the~\citet{kennicutt98} H$\alpha$-SFR relationship. This relationship is based on an assumption of constant SFR at ages $<2\times10^7$ years. The fact that there is general agreement with UV SFRs estimated using similar assumptions suggests that a recent, constant SFR model provides an adequate description of our targets. 


\subsection{IR}

Infrared fluxes were gathered from the literature for 12 of our targets, eight from Spitzer data and four from IRAS (Table~\ref{table7}).  Spitzer fluxes were used to determine total IR luminosity following the method of~\citet{quillen08}, who interpolate a $15\mu$m flux from $8\mu$m and $24\mu$m Spitzer data and then employ the relationship $L_{\rm{IR}}=(11.1^{+5.5}_{-3.7})\times(\nu L_{\nu}[15~\mu{\rm{m}}])^{0.998}$~\citep{elbaz02}. We note that this relationship will over-estimate the IR luminosity from star formation if the $24\mu{\rm{m}}$ point is contaminated with AGN emission. For the four objects without Spitzer data, IRAS fluxes were converted to total IR luminosity using $L_{\rm{IR}}\sim1.7L_{60~\mu{\rm{m}}}$~\citep{rowan97}. 

Total IR luminosities were then converted to SFRs using equation (5) of~\citet{bell03} (as in~\citealt{odea08}) 

\begin{equation}
\psi (M_\odot~{\rm{yr}}^{-1} ) = A \left({{L_{\rm{IR}}}\over{L_\odot}}\right)(1+\sqrt{10^9 L_\odot / L_{\rm{IR}}})
\end{equation}

\noindent where $A=1.17 \times 10^{-10}$ for $L_{\rm{IR}}<10^{11} L_\odot$ and $A=1.57\times 10^{-10}$ at higher luminosities. These relationships were constructed using a sample of galaxies that did not include any early-types and therefore may not be well suited to our target population. In addition there is significant expected scatter~\citep[at least 50\% at $10^9$ L$_\sun$ and 25\% at $10^{11}$ L$_\sun$][]{bell03}.

Overall, IR inferred SFRs are a factor of $\sim10$ higher than UV inferred (constant) SFRs, though our results are somewhat more consistent with recent Herschel estimates~\citep[Abell 2597 and ZwCl 3146;][]{edge10}. Because the IR luminosities are so large and are presumably due to dust absorbing UV radiation, the larger inferred IR SFRs potentially indicate high UV extinction. To further explore the attenuation in our targets, we calculate their IR excesses~\citep[e.g.,][]{gordon00}: 

\begin{equation}
{\rm{IRX}} ={\rm{log}}_{10}~(L_{\rm{dust}}/L_{\rm{UV}})
\end{equation}
\noindent
where $L_{\rm{UV}}=\nu L_{\nu}$ and $\nu=c/1390$~\AA.

We compare the IRX and FUV-NUV colors of our sample with~\citet{johnson07}, who investigate these properties in SDSS galaxies (Figure~\ref{fig9}). Overall the colors of the BCGs in our sample fall into a region that is heavily populated with the colors of galaxies exhibiting recent star formation.  Many of our targets have high values of IRX, consistent with heavily dust-enshrouded star formation which is very common in rapidly star forming galaxies.

\subsection{X-ray}


Cluster entropy can be used as a tool for investigating the energy budget of baryons in clusters~\citep[e.g.,][]{ponman99}. Thermodynamic entropy is proportional to the log of the measurable quantity $K \equiv T_{\rm{X}} / n_{\rm{e}}^{2/3}$. Likewise, the time it takes for intracluster gas to radiatively cool is also proportional to its density and temperature: $t_{cool}\propto T^{1/2} n^{-1}$. In general, if the cooling time is shorter than $\sim10^9$ years, clusters tend to exhibit cool core characteristics~\citep{hudson09}. Our sample was chosen to span a range of cooling times so that we could examine the relationship between these parameters and UV emission. 

The X-ray properties of our sample are taken from the ACCEPT database\footnote{\url{http://www.pa.msu.edu/astro/MC2/accept/}}~\citep{cavagnolo09}.  As well as using central estimates based on a fit, both entropy (K) and cooling time profiles were interpolated to obtain values at R=20 kpc from the center of each cluster. Interpolation enabled a more uniform comparison between objects with different redshifts and/or data quality.

NUV inferred SFR vs. entropy is shown in Figure~\ref{fig10}, and SFR is plotted vs. cluster cooling time in Figure~\ref{fig11}. The plots constructed with central (R$\rightarrow0$) entropies and cooling times indicate thresholds comparable to those reported by~\citet{cavagnolo08b},~\citet{voit08}, and~\citet{rafferty08}. Comparisons with gas properties at R $=20$ kpc from the cluster center, however, yield smoother trends in entropy and cooling time. These plots show a clear tendency for lower entropy, shorter cooling time objects to exhibit more star formation, providing convincing evidence that the star formation in these objects is directly related to cooling gas in the cluster cores. A BCES regress fit between NUV inferred SFR and the R $=20$ kpc cooling time data yields the relationship ${\rm{log_{10}}}~{\rm{SFR}_{\rm{NUV}}} = a~{\rm{log_{10}}}~{\rm{t}_{\rm{cool, 20}}} + b $, where $a=- 3.9\pm{0.7}$ and $b=- 0.6\pm{0.1}$.

\section{Summary and Discussion \label{discussion}}

In our UV study of 17 cool core clusters we find that GALEX easily detects star formation in cluster BCGs out to $z\ge 0.45$ and to unprecedented radii. The BCGs are found to be bluest in the center, with colors that become increasingly redder with radius, suggesting that star formation is most easily detected in the central regions.

We construct UV/J band calibration relationships from 17 cluster ellipticals and 22 quiescent BCGs that enable us to subtract the expected UV light from older populations.  In most of the CC clusters studied, we find significant UV luminosity excesses and colors that strongly suggest recent and/or current star formation. 

Star formation rates are estimated using Starburst99 templates for both continuous and burst models, for easy comparison to results in the literature from other wavebands. Our findings are corroborated by H$\alpha$ observations, showing good agreement with~\citet{kennicutt98} models of recent, continuous star formation. 

To investigate attenuation in CC BCGs, IRX values are calculated using our GALEX data and IR fluxes from the literature. Comparisons with the SDSS sample of~\citet{johnson07} indicate that our sample has moderate-to-high extinction and has NUV-FUV colors consistent with or bluer than other star-forming galaxies. These results emphasize a need for additional observations and detailed studies of cluster BCGs, as currently there are no adequately large samples of quiescent BCGs to provide a sufficient context for our findings.

We also compare our UV results to properties of the intragalactic medium using X-ray observations. We find clear correlations between UV excess, cluster entropy, and central cooling time, demonstrating that the star formation is directly and incontrovertibly related to the cooling gas in these objects.


\acknowledgements

Support for this work was provided by NASA through GALEX award NNX07AJ38G, Chandra Archive award AR7-8012X, and LTSA grant NNG05GD82G. This research has made use of the NASA/IPAC Extragalactic Database (NED) which is operated by the Jet Propulsion Laboratory, California Institute of Technology, under contract with the National Aeronautics and Space Administration.












\begin{figure}
\centerline{\includegraphics[width=6.5in]{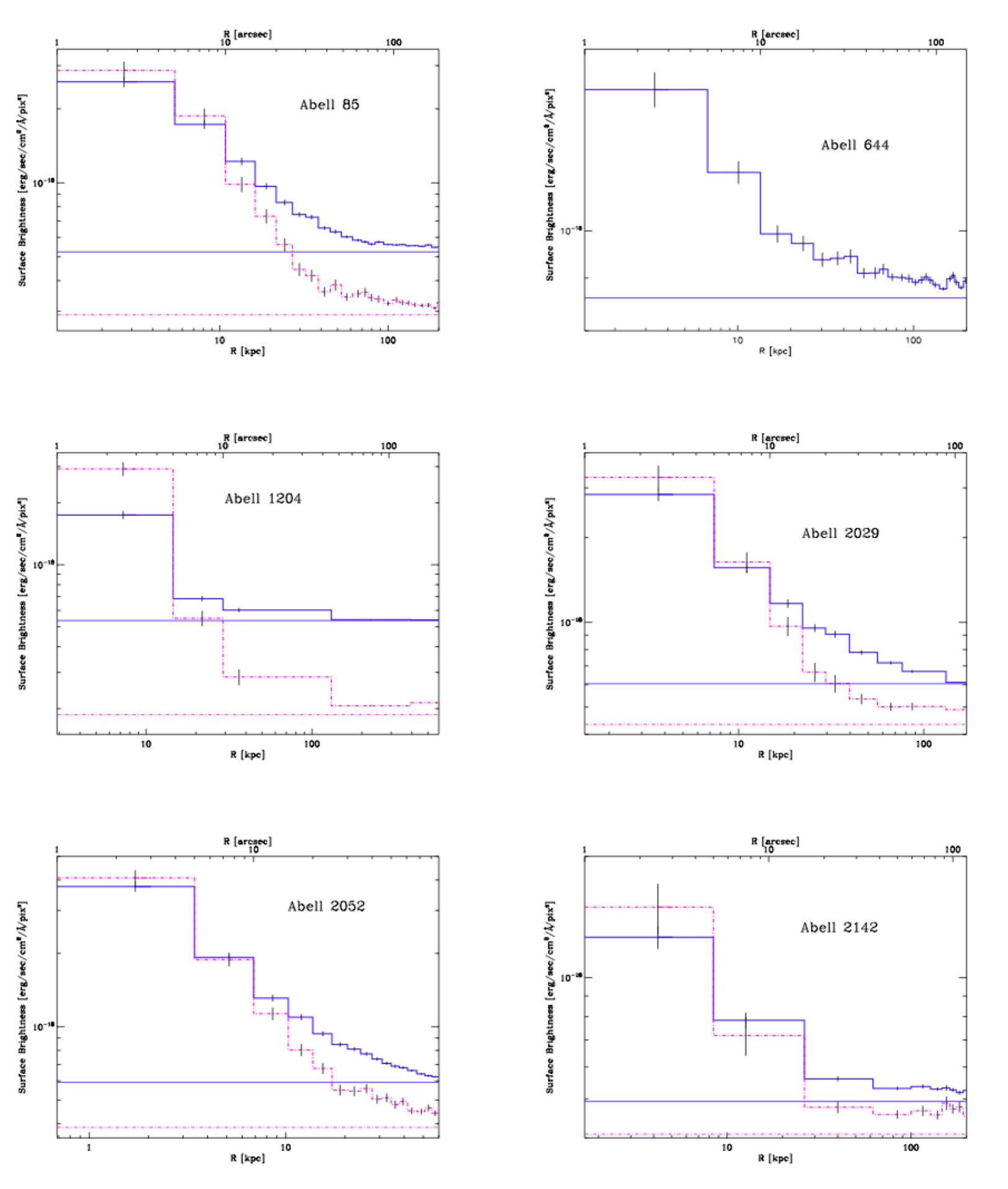}}
\end{figure}
\begin{figure}
\centerline{\includegraphics[width=6.5in]{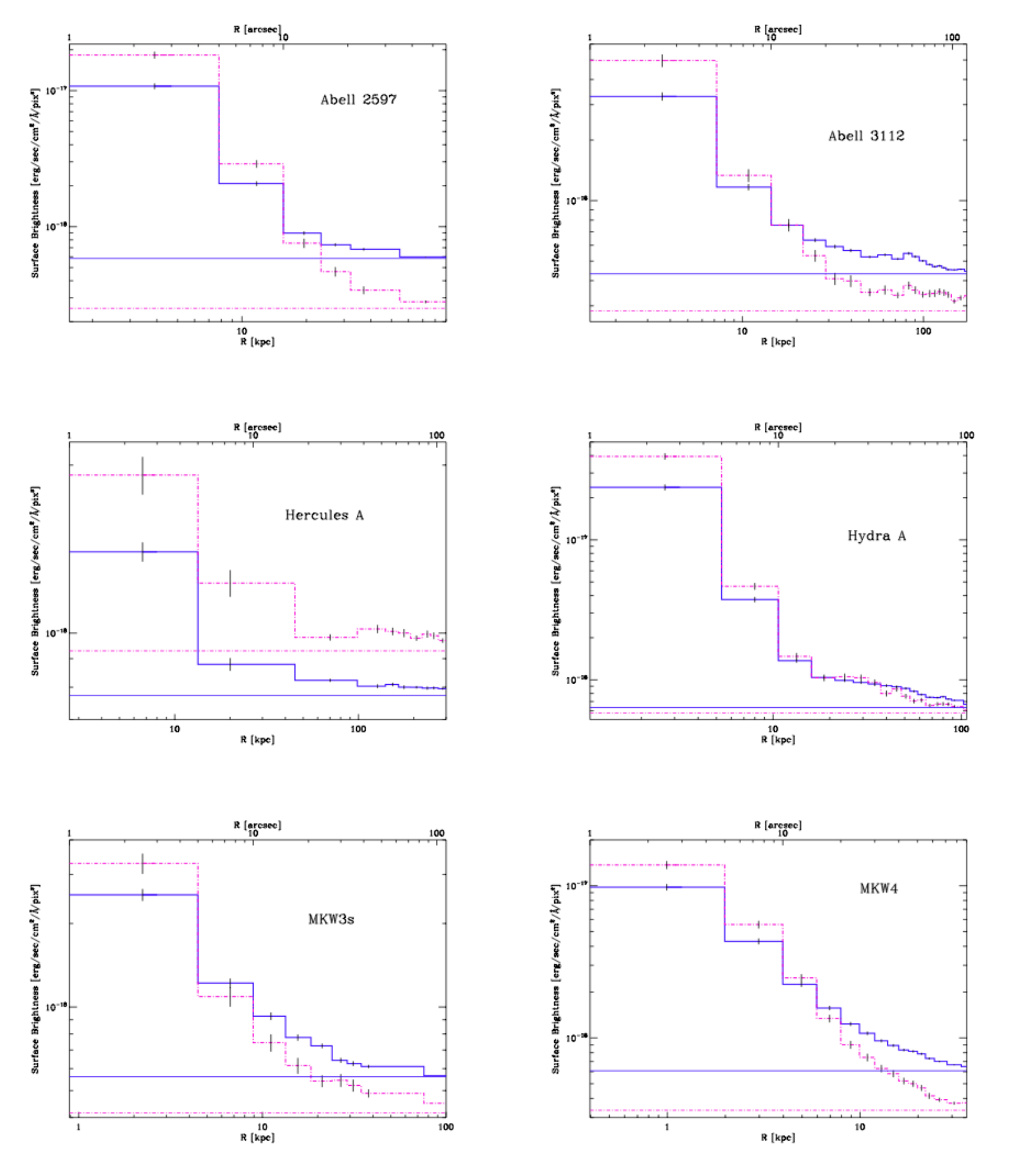}}
\end{figure}
\begin{figure}
\centerline{\includegraphics[width=6.5in]{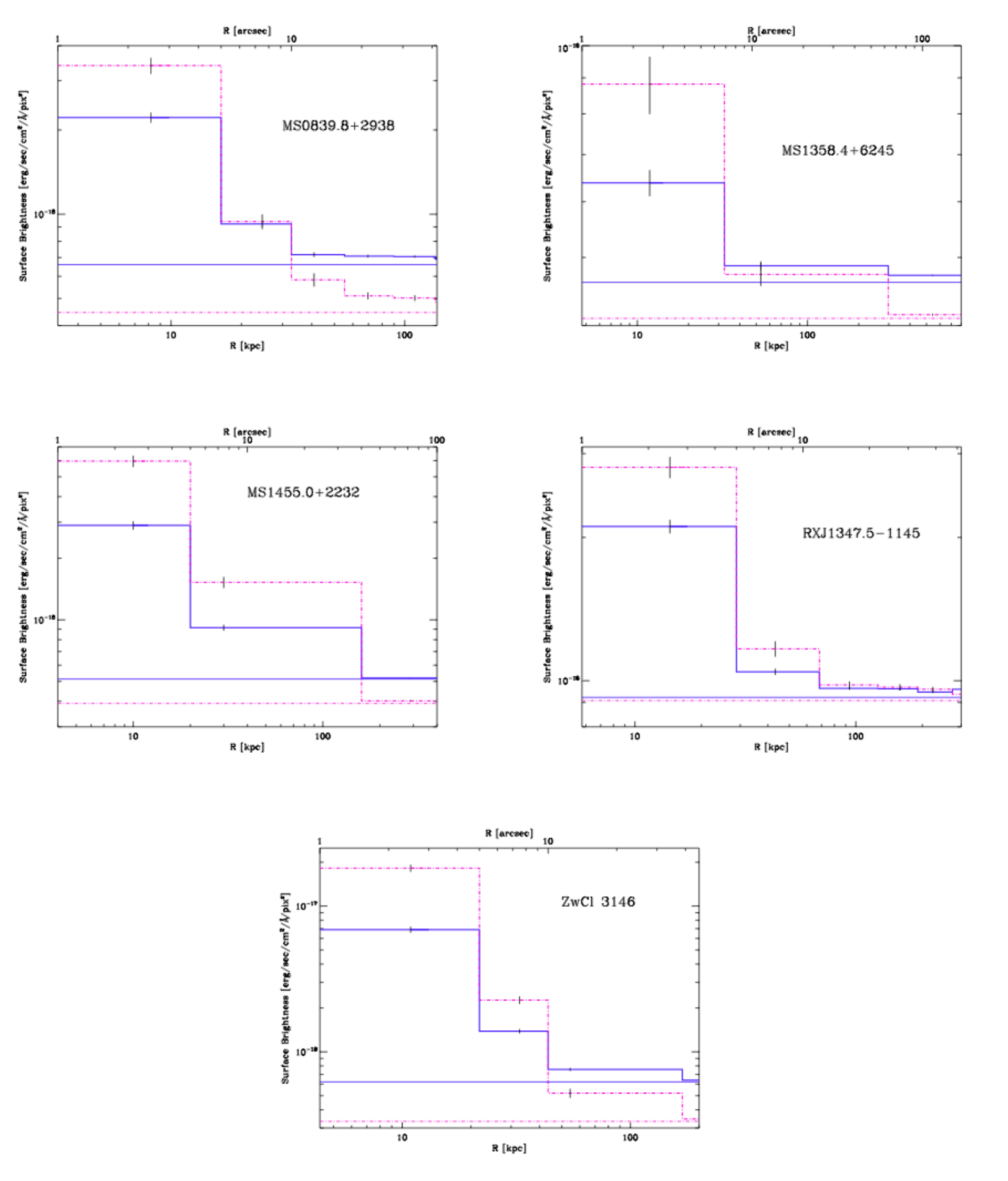}}
\caption{UV surface brightness profiles for the CC clusters in our sample. The NUV profiles are in blue (solid lines) and the FUV profiles are in pink (dot-dash). Horizontal lines indicate background levels. Profiles were constructed to reach S/N $> 3$ in each bin. Note that UV emission is often detectable out to $>100$ kpc with GALEX. \label{fig2}}
\end{figure}

\begin{figure}
\centerline{\includegraphics[width=6.5in]{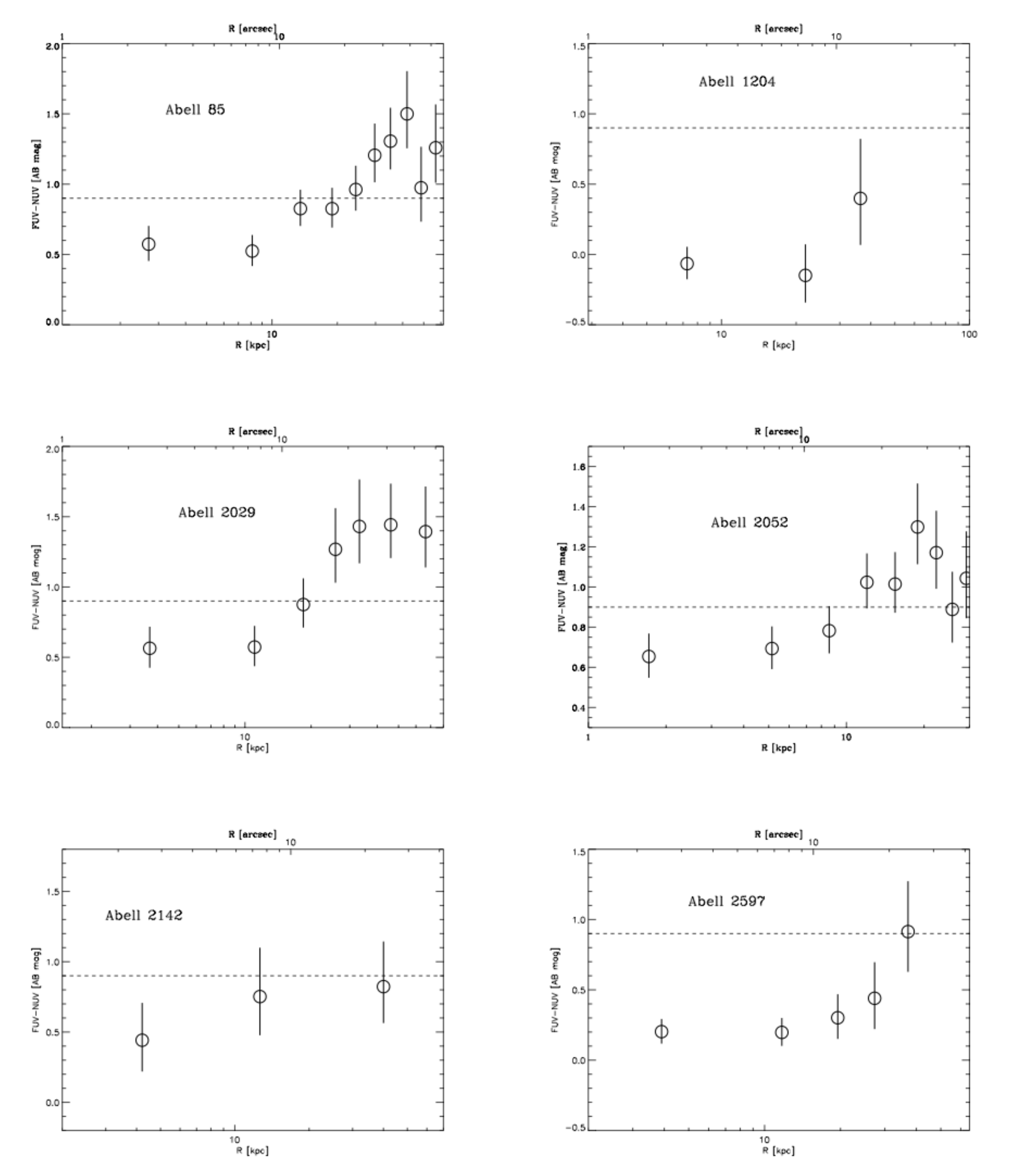}}
\end{figure}
\begin{figure}
\centerline{\includegraphics[width=6.5in]{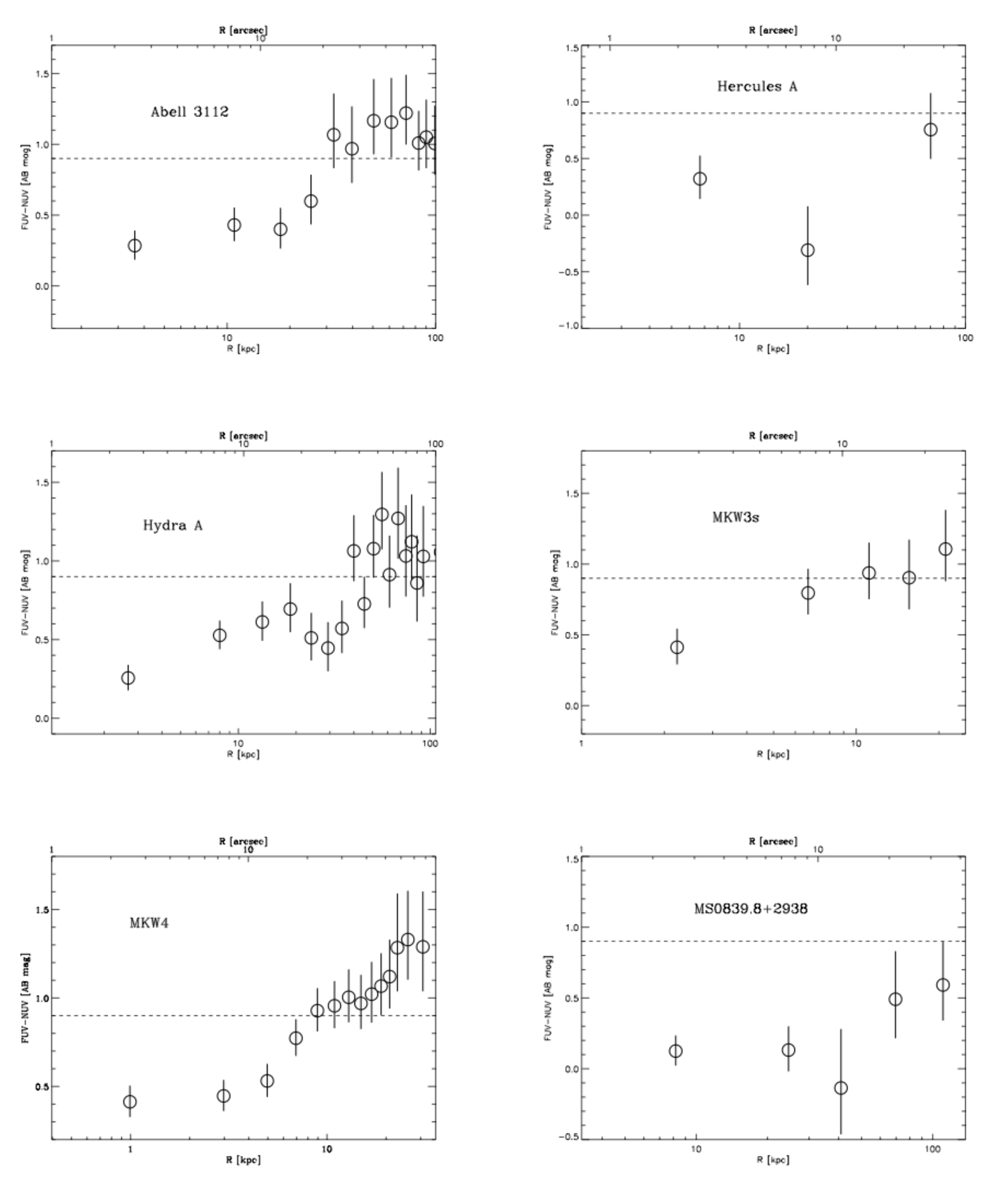}}
\end{figure}
\begin{figure}
\centerline{\includegraphics[width=6.5in]{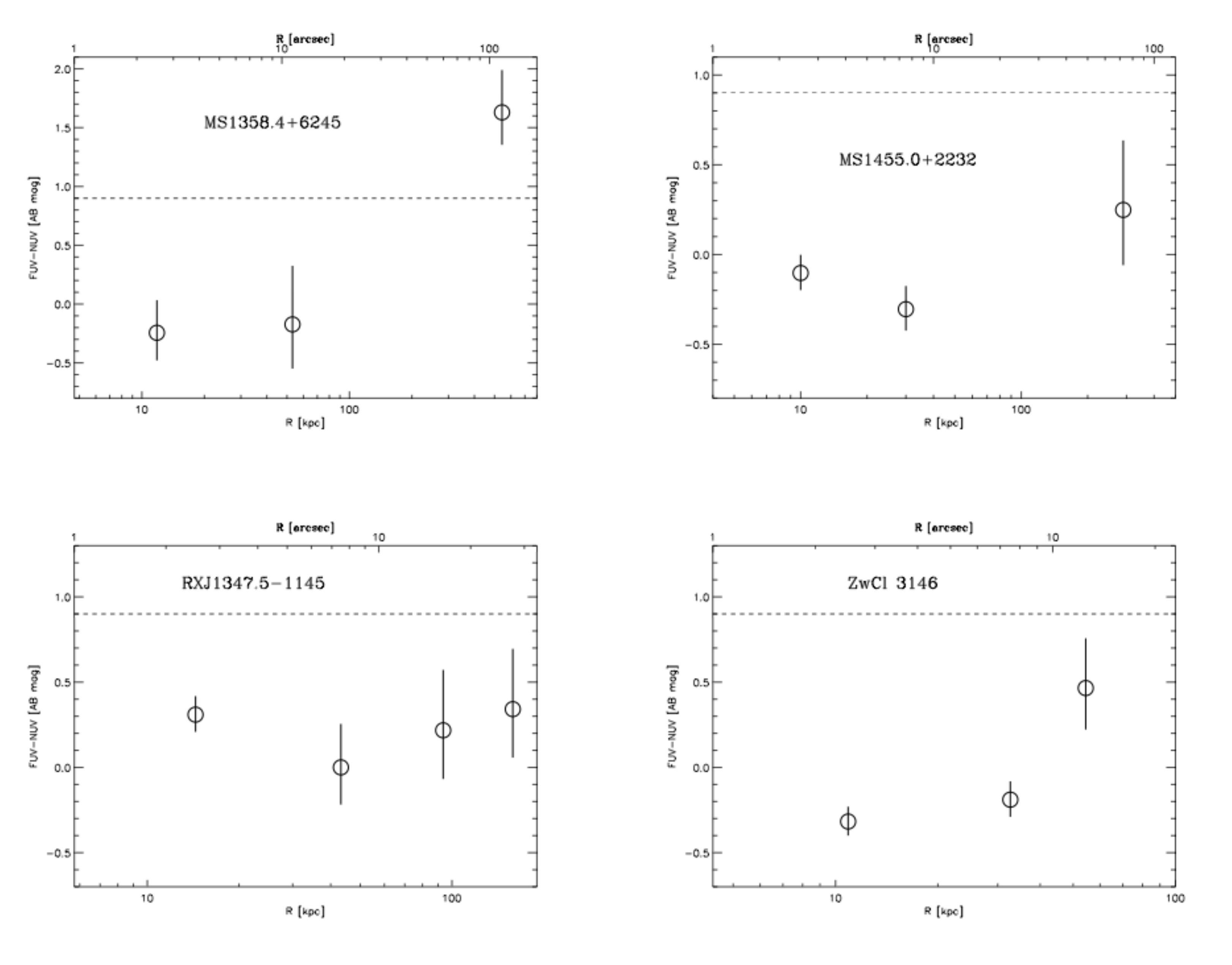}}
\caption{UV color profiles (FUV-NUV), clearly showing bluer emission in the center. Star forming galaxies tend to have GALEX UV colors $\sim0.4$, while the majority ($82\%$) of elliptical galaxies have UV colors of $>0.9$ \citep{gildepaz07}. Dotted lines are shown at FUV-NUV=0.9 as a point of reference.\label{fig3}}
\end{figure}

\begin{figure}
\centerline{\includegraphics[width=6.5in]{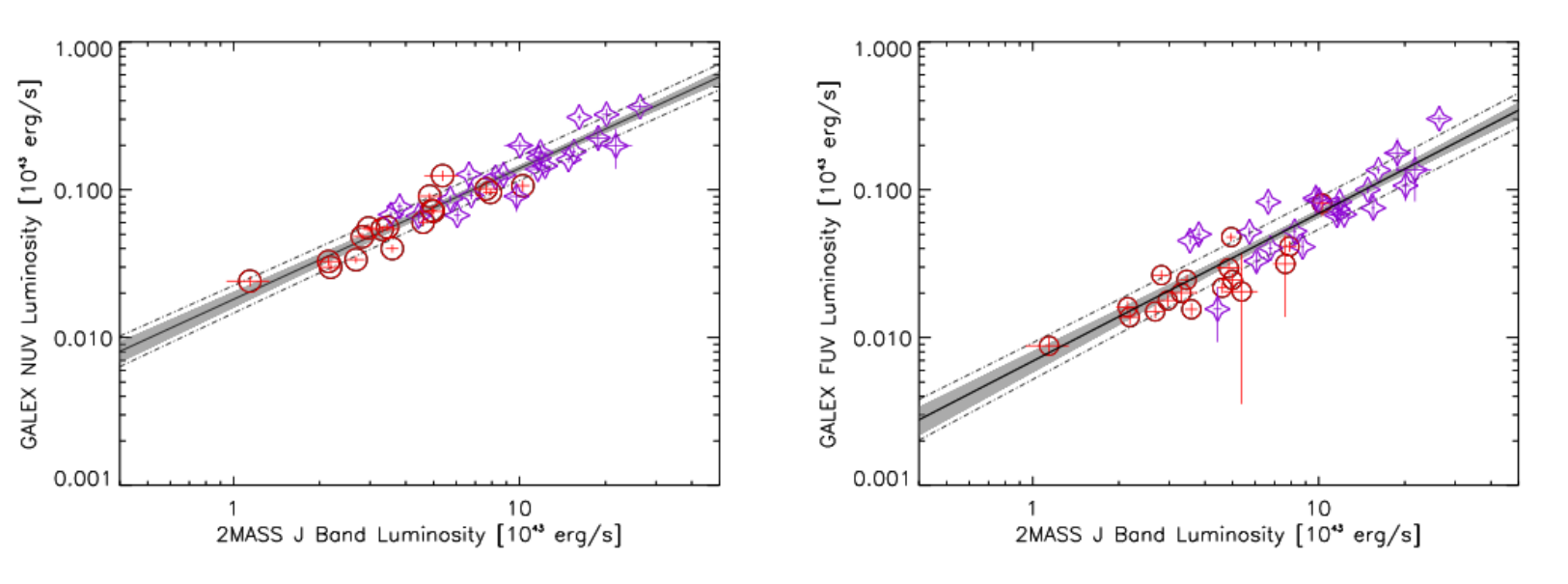}}
\caption{{\bf Calibration Relationships.} UV vs. J band relationships obtained from our 39 passive (non-star forming) calibration targets (17 cluster ellipticals, 22 BCGs). Cluster ellipticals are shown as circles, and BCGs as 4-pointed stars. Shaded areas represent 1$\sigma$ errors on the relationship, while dot-dash lines indicate the scatter ($\sim20\%$ in NUV and $\sim30\%$ in FUV).\label{fig4}}
\end{figure}
 
\begin{figure}
\centerline{\includegraphics[width=6.5in]{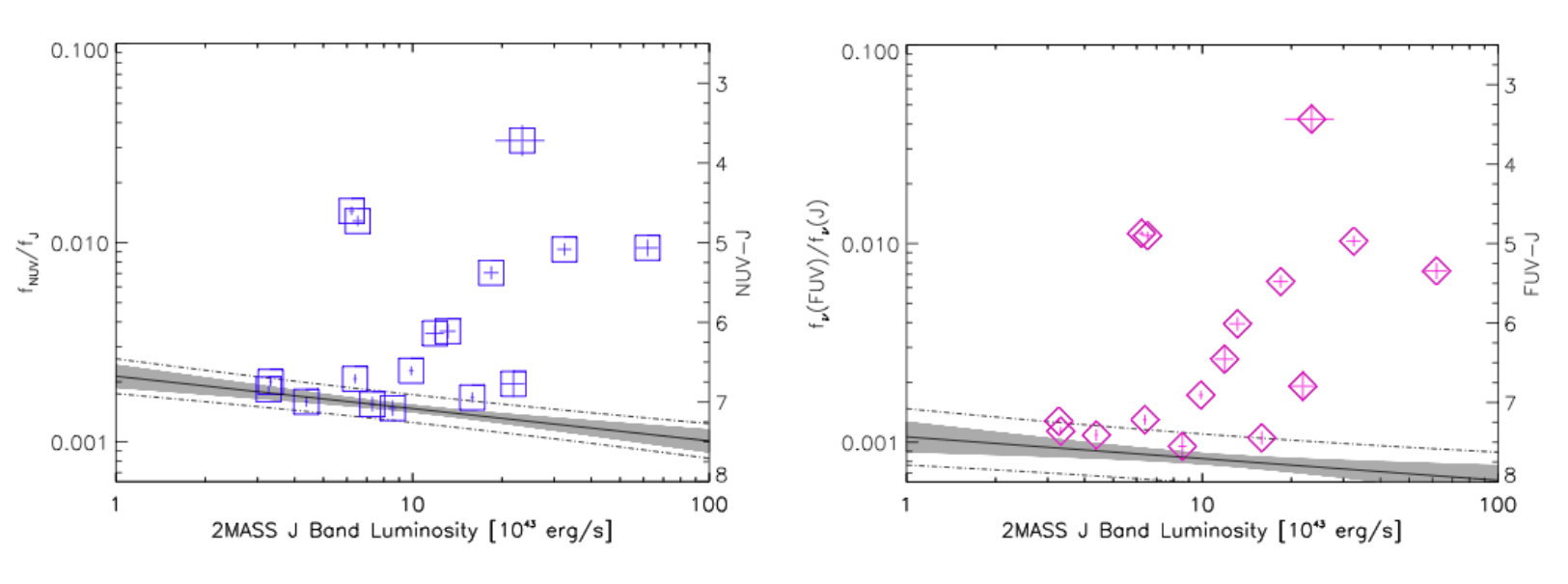}}
\caption{UV/J flux ratios for our CC sample. Lines show the fitted relationships for passively evolving galaxies from our correlations; shaded regions designate $1\sigma$ errors on the fits. Dot-dash lines indicate the scatter. Most of the objects in our sample exhibit signs of recent star formation. \label{fig5}}
\end{figure}

\begin{figure}
\centerline{\includegraphics[width=8.in]{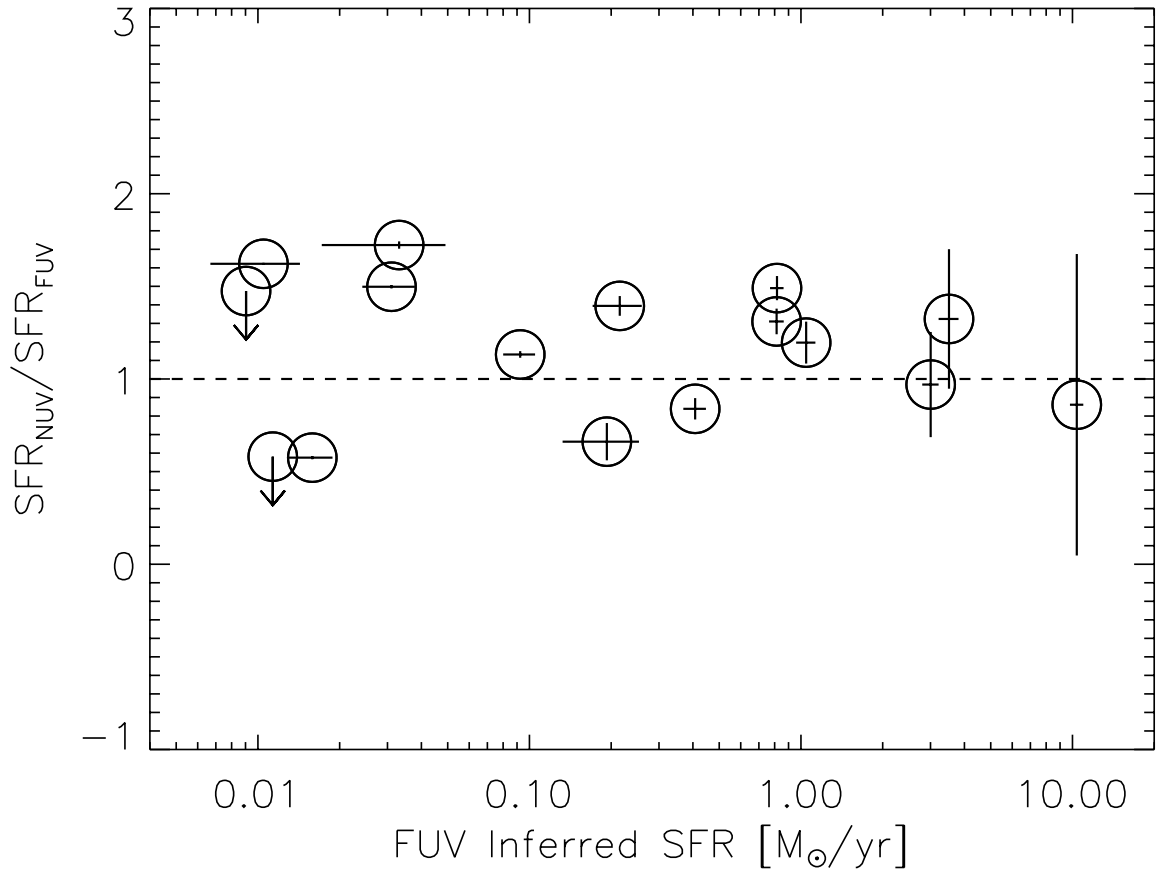}}
\caption{Ratio of NUV to FUV inferred star formation rates vs. FUV SFR. The average ratio is 1.2, suggesting that higher scatter in the FUV calibration relationship may result in a slight underestimation of star formation rates. \label{fig6}}
\end{figure}
  
\begin{figure}
\centerline{\includegraphics[width=6.in,angle=90]{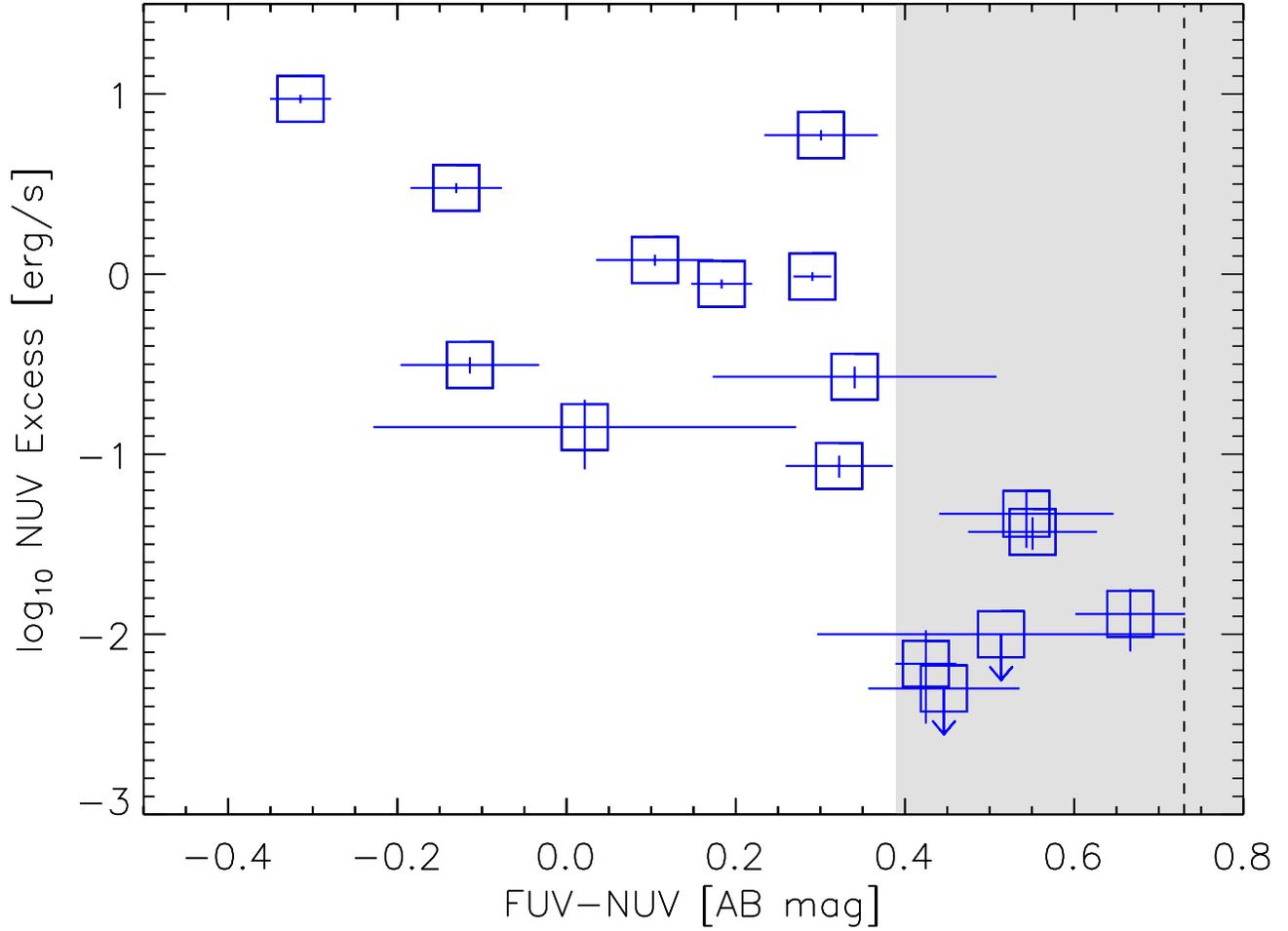}}
\caption{Total 7\arcsec FUV-NUV color vs. NUV luminosity excess. The dashed line shows the mean color of our control sample (0.73), while the shaded area designates the standard deviation (0.34). The object with high NUV excess at FUV-NUV$\sim0.3$ is suspected of having some contaminating Ly$\alpha$ emission contributing to its NUV flux. \label{fig7}}
\end{figure}

\begin{figure}
\centerline{\includegraphics[width=8in,angle=180]{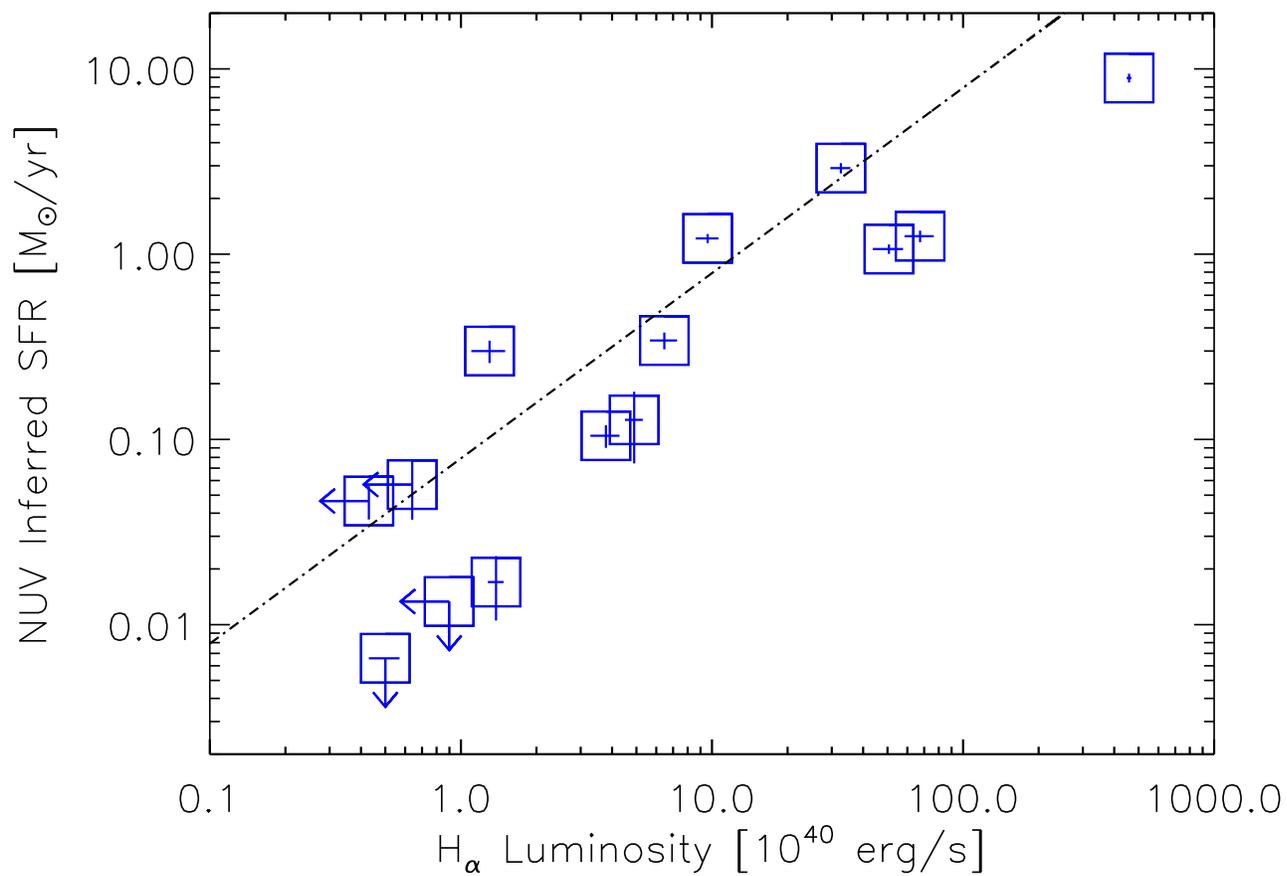}}
  \caption{NUV inferred SFR vs. H$\alpha$ luminosity for a subset of our targets. The line indicates the L$_{\rm{H}\alpha}$-SFR relationship of \citet{kennicutt98}. The general agreement between the line and our data suggests that a 20 Myr constant star formation model provides a reasonable description of our targets. \label{fig8}}
\end{figure}

\begin{figure}
\centerline{\includegraphics[width=6.5in]{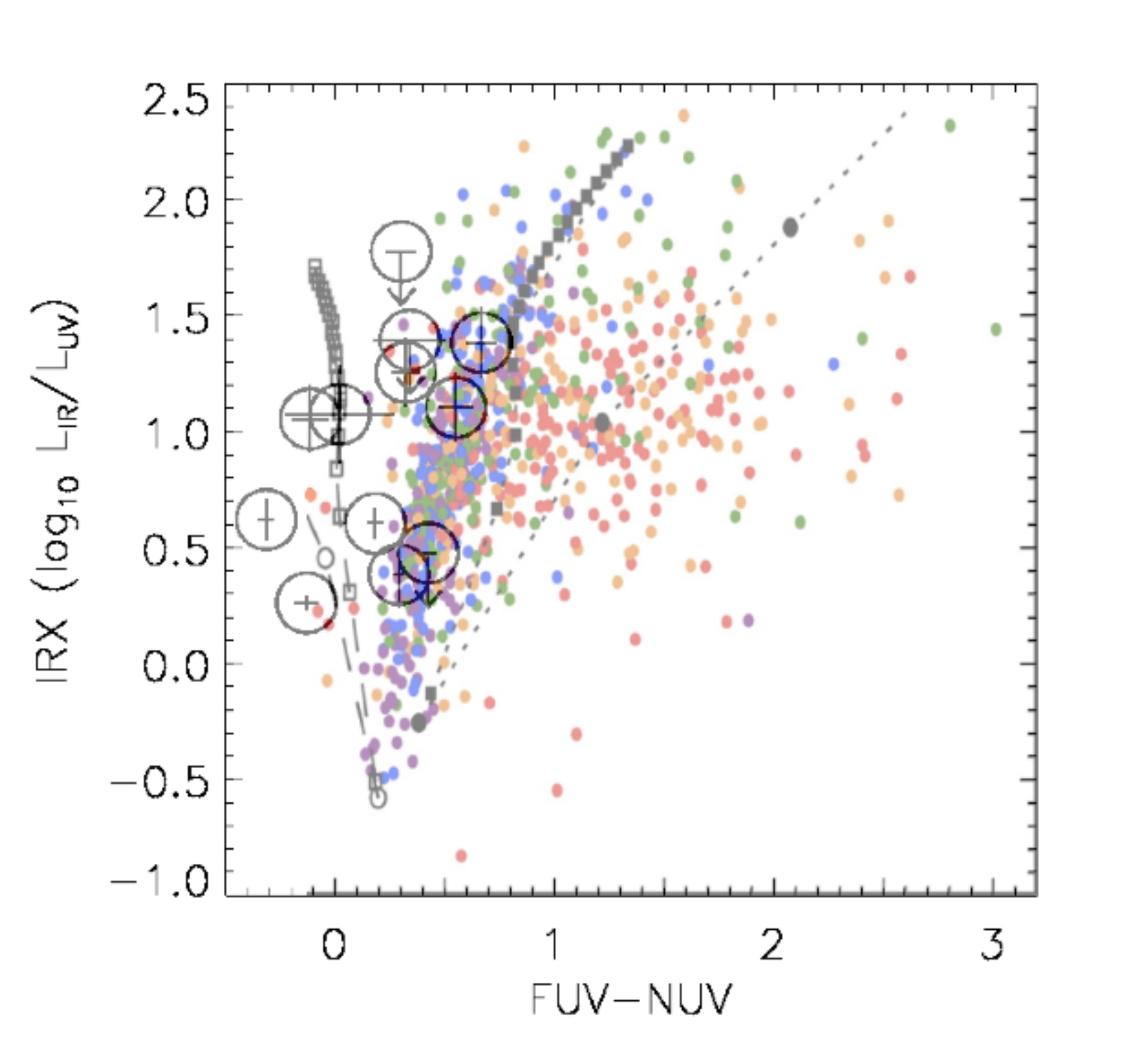}}
  \caption{IR excess (IRX) vs. UV color for our sample of CC BCGs (large circles) plotted over Figure 1 of~\citet{johnson07}. Small black symbols with dashed lines show the effects of dust attenuation for MW (open symbols) and SMC (filled symbols) extinction laws~\citep{witt00}. Colors indicate values of Dn(4000), the strength of the 4000~\AA~break, from purple ($\sim1$) through red ($\sim2$), with lower values suggesting more recent star formation. The positions of our targets on this plot are consistent with galaxies exhibiting dusty star formation.
  \label{fig9}}
\end{figure}

\begin{figure}
\centerline{\includegraphics[width=8.in]{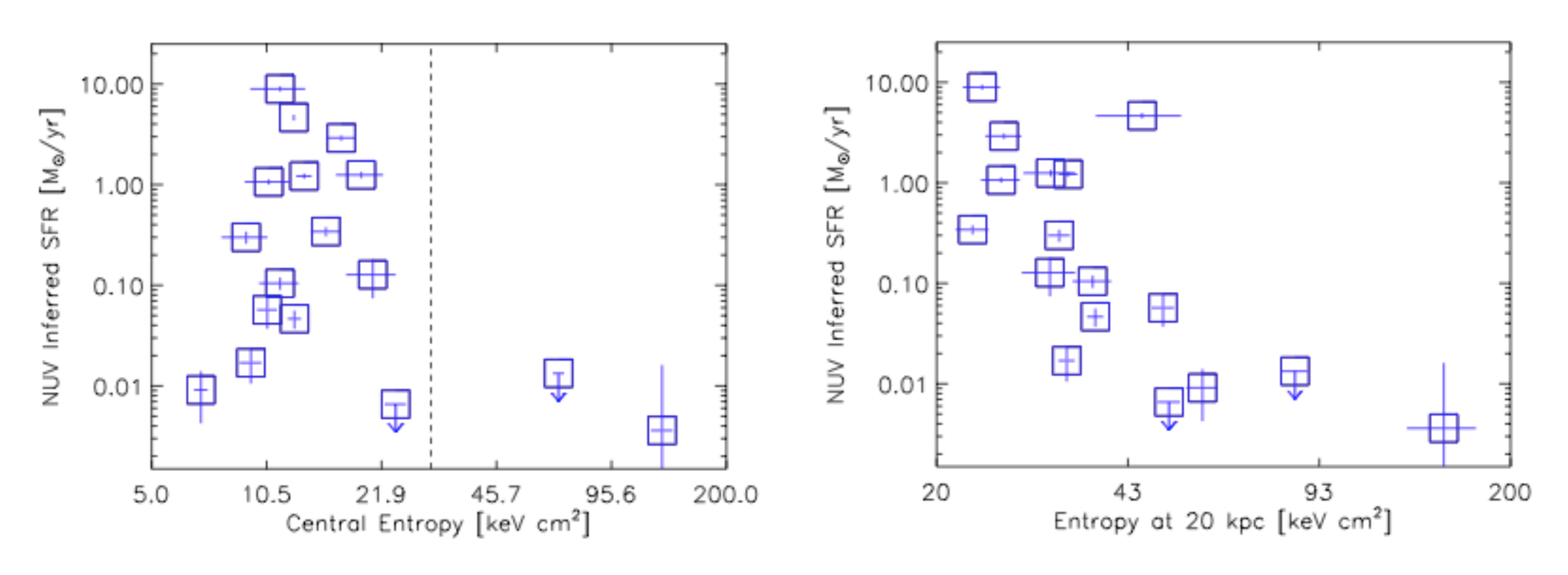}}
\caption{{\it{Left:}} NUV inferred SFR vs. central gas entropy. A dashed line indicates the value of the entropy threshold reported by~\citet{cavagnolo08b}. {\it{Right:}} SFR vs. entropy measured at R$=20$ kpc, showing a tendency for more star formation to occur in lower entropy objects. The outlying point is our highest-$z$ cluster, for which 20 kpc is only 3.5\arcsec, so it is likely unresolved in the X-ray profile at that radius. Based on its cooling time, it should have an R $=20$ kpc entropy closer to 25 keV cm$^2$.\label{fig10}} 
\end{figure}

\begin{figure}
\centerline{\includegraphics[width=8.in]{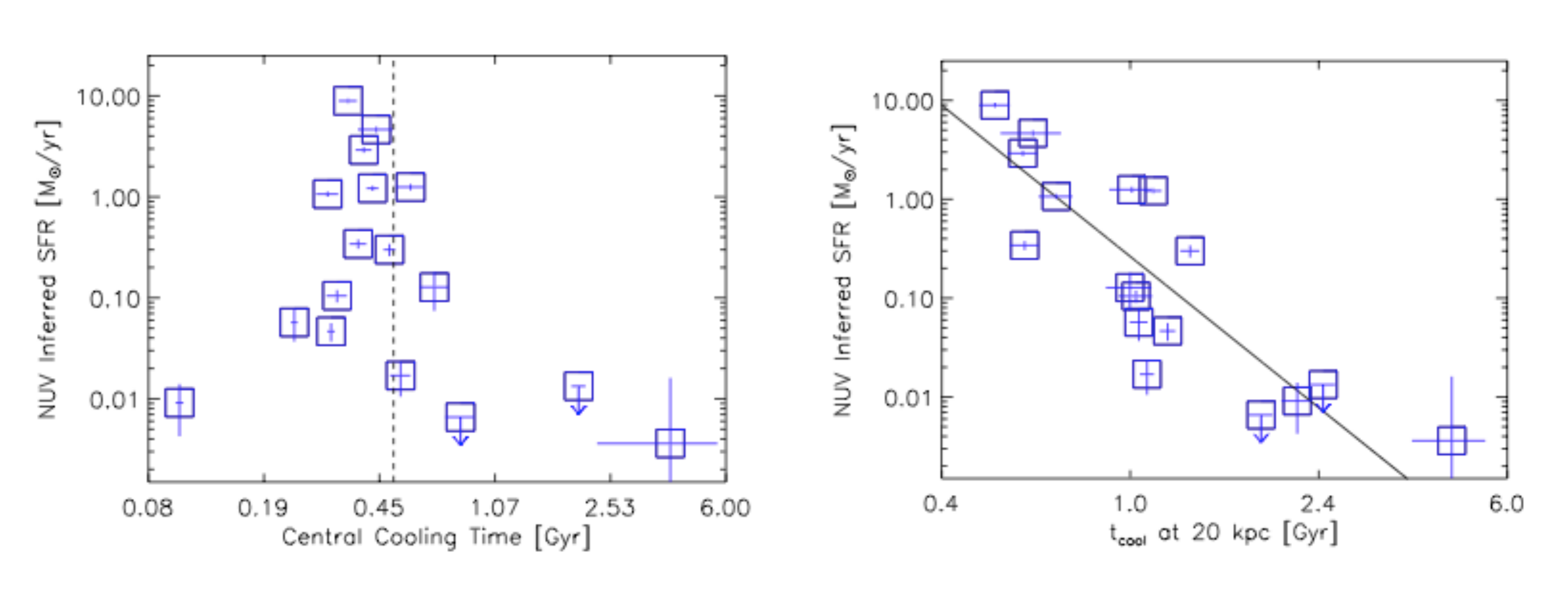}}
  \caption{{\it{Left:}} NUV inferred SFR vs. central cooling time. Here the dashed line corresponds to the observed cooling time threshold of~\citet{rafferty08}. The object with low SFR and very short cooling time is our lowest-$z$ cluster, which was much more highly resolved than the rest of the sample (down to the inner 2 kpc). {\it{Right:}} SFR vs. cooling time measured at R$=20$ kpc, with best fitting powerlaw overlaid. This plot clearly suggests that a relationship exists between cooling plasma and star formation in cool core cluster BCGs.\label{fig11}}
\end{figure}


		

\clearpage
\begin{deluxetable}{lcc}
\tablecolumns{3}
\tablewidth{0pt}
\tablecaption{GALEX Cool Core Sample\label{table1}}
\tablehead{
\multicolumn{1}{l}{Cluster}
  & \multicolumn{1}{c}{$z$}
  & \multicolumn{1}{c}{Exposure} \\
  \multicolumn{1}{l}{}
  & \multicolumn{1}{c}{}
  & \multicolumn{1}{c}{(NUV/FUV) [s]}
}
\startdata
Abell 85 & 0.0557 & 2494 / 2494 \\
Abell 644 & 0.0705 & 3520 / 0\tablenotemark{a} \\
Abell 1204 & 0.1706 & 3738 / 3738 \\
Abell 2029 & 0.0779 & 1517 / 1517 \\
Abell 2052 & 0.0345 & 2863 / 2863 \\
Abell 2142 & 0.0904 & 1556 / 1556 \\
Abell 2597 & 0.0830 & 2111 / 2111 \\
Abell 3112 & 0.0761 & 4873 / 2618 \\
Hercules A & 0.1540 & 3870 / 3870 \\
Hydra A   & 0.0549 & 2230 / 2230\\
MKW3s   & 0.0453 & 2271 / 2271\\
MKW4   & 0.0196& 2194 / 2194\\
MS0839.8+2938   & 0.1980 & 4729 / 4728\\
MS1358.4+6245  & 0.3272 & 5614 / 5614\\
MS1455.0+2232  & 0.2578 & 3385 / 3384\\
RXJ1347.5-1145  & 0.4500 & 9120 / 9119\\
ZwCl 3146  & 0.2906 & 3127 / 3127\\
\enddata 
\tablenotetext{a}{Observation executed after loss of FUV detector.}
\end{deluxetable}

\clearpage
\begin{deluxetable}{lllccc}
\tablecolumns{6}
\tablewidth{0pt}
\tablecaption{Brightest Cluster Galaxy 7\arcsec Radius Photometry\label{table2}}
\tablehead{
\multicolumn{1}{l}{Cluster}
&\multicolumn{2}{c}{Position}
  & \multicolumn{3}{c}{{mAB}}\\
 \multicolumn{1}{l}{}
&\multicolumn{1}{c}{RA}
& \multicolumn{1}{c}{Dec}
  & \multicolumn{1}{c}{\small{FUV}}
  & \multicolumn{1}{c}{\small{NUV}}  
  & \multicolumn{1}{c}{\small{J}\tablenotemark{a}}
  }
\startdata
Abell 85 &  00 41 50.4 & -09 18 11.0 &${20.86}^{+0.07}_{-0.07}$ & ${20.31}^{+0.03}_{-0.03}$ & ${13.71}^{+0.01}_{-0.01}$\\
Abell 644 &  08 17 25.6 & -07 30 45.6 &\nodata & ${20.94}^{+0.08}_{-0.07}$ & ${14.10}^{+0.03}_{-0.03}$\\
Abell 1204 & 11 13 20.5 & +17 35 41.5 &${21.35}^{+0.07}_{-0.06}$ & ${21.47}^{+0.05}_{-0.05}$ & ${15.53}^{+0.07}_{-0.07}$\\
Abell 2029 & 15 10 56.1 & +05 44 40.3&${20.88}^{+0.1}_{-0.09}$ & ${20.34}^{+0.04}_{-0.04}$ & ${13.49}^{+0.01}_{-0.01}$ \\
Abell 2052 & 15 16 44.5 & +07 01 17.8& ${20.65}^{+0.06}_{-0.06}$ & ${19.98}^{+0.03}_{-0.02}$ & ${13.35}^{+0.01}_{-0.01}$\\
Abell 2142 & 15 58 20.0 & +27 14 00.2& ${22.0}^{+0.2}_{-0.2}$ & ${21.50}^{+0.09}_{-0.08}$ & ${14.50}^{+0.04}_{-0.03}$ \\
Abell 2597 &23 25 19.7&-12 07 27.2& ${19.21}^{+0.03}_{-0.03}$ & ${19.03}^{+0.02}_{-0.02}$ & ${14.60}^{+0.04}_{-0.04}$ \\
Abell 3112 & 03 17 57.6 & -44 14 17.2 &${20.75}^{+0.06}_{-0.06}$ & ${20.43}^{+0.02}_{-0.02}$ & ${13.95}^{+0.02}_{-0.02}$ \\
Hercules A & 16 51 08.2 & +04 59 33.9 &${21.7}^{+0.2}_{-0.1}$ & ${21.36}^{+0.08}_{-0.07}$ & ${15.39}^{+0.08}_{-0.08}$ \\
Hydra A   & 09 18 05.7 & -12 05 44.2 &${18.29}^{+0.02}_{-0.02}$ & ${18.00}^{+0.01}_{-0.01}$ & ${13.71}^{+0.02}_{-0.02}$\\
MKW3s   &15 21 51.8& +07 42 32.5&${21.01}^{+0.08}_{-0.08}$ & ${20.57}^{+0.04}_{-0.04}$ & ${13.66}^{+0.02}_{-0.01}$\\
MKW4   & 12 04 27.1 & +01 53 45.9 &${19.28}^{+0.03}_{-0.03}$ & ${18.85}^{+0.02}_{-0.02}$ & ${12.116}^{+0.005}_{-0.005}$\\
MS0839.8+2938   & 08 42 56.0 & +29 27 26.8 &${20.77}^{+0.06}_{-0.06}$ & ${20.66}^{+0.04}_{-0.03}$ & ${15.52}^{+0.06}_{-0.06}$ \\
MS1358.4+6245  & 13 59 50.5 & +62 31 04.2 & ${23.3}^{+0.2}_{-0.2}$ & ${23.2}^{+0.2}_{-0.1}$ & ${16.6}^{+0.1}_{-0.1}$\\
MS1455.0+2232  &14 57 15.1 & +22 20 33.1& ${20.24}^{+0.05}_{-0.04}$ & ${20.37}^{+0.03}_{-0.03}$ & ${15.55}^{+0.06}_{-0.06}$\\
RXJ1347.5-1145  &13 47 30.7 & -11 45 09.3&${21.36}^{+0.06}_{-0.06}$ & ${21.06}^{+0.03}_{-0.03}$ & ${16.26}^{+0.1}_{-0.09}$\\
ZwCl 3146  & 10 23 39.6 & +04 11 11.6 &${19.22}^{+0.03}_{-0.03}$ & ${19.53}^{+0.02}_{-0.02}$ & ${16.2}^{+0.2}_{-0.2}$\\
\enddata 
\tablenotetext{a}{From the 2MASS Database.}
\end{deluxetable}

\clearpage
\begin{deluxetable}{lcc}
\tablecolumns{3}
\tablewidth{0pt}
\tablecaption{Control Sample\label{table3}}
\tablehead{
\multicolumn{1}{l}{Target}
  & \multicolumn{1}{c}{$z$}
  & \multicolumn{1}{c}{GALEX Exposure} \\
  \multicolumn{1}{l}{}
  & \multicolumn{1}{c}{}
  & \multicolumn{1}{c}{(NUV/FUV) [s]}
}
\startdata
\multicolumn{3}{c}{\underline{Cluster Ellipticals}} \\   
Abell 389\tablenotemark{a}  & 0.1133 & 31253 / 21733 \\
Abell 795  & 0.1359 & 2888 / 2888 \\
Abell 2100  & 0.1533 & 3446 / 1739 \\
Abell 2670\tablenotemark{a} & 0.0762 & 21700 / 21693 \\
\multicolumn{3}{c}{\underline{BCGs}} \\   
Abell 119  &  0.0442 & 3013 / 3013\\
Abell 951  & 0.143 & 35392 / 17471 \\
Abell 963 & 0.206 & 29040 / 29040 \\
Abell 1278  &  0.129 & 1616 / 1616\\
Abell 1367  &  0.0220 & 2876 / 2876 \\
Abell 1882  &  0.139 & 1820 / 1820\\
Abell 2218  &  0.1720 & 12220 / 7186\\
Abell 2219  &  0.2244 & 2742 / 2742\\
Abell 2235  &  0.151 & 41915 / 29079\\
Abell 2240  &  0.137 & 1583 / 1582\\
Abell 2249 &  0.079 & 2678 / 2678 \\
Abell 2255  & 0.0805 & 1649 / 1649 \\
Abell 2399  &  0.058 & 2973 / 2973\\
Abell 2448  &  0.083 & 6425 / 1658\\
Abell 2631  &  0.2772 & 3374 / 3374\\
Abell 3158  &  0.0580 & 1024 / 1024\\
Abell 3330  &  0.0918 & 59982 / 22666 \\
Abell 3376  &  0.0456 & 2132 / 2132 \\
Abell 3556  &  0.048 & 1688 / 1688\\
Abell 3716  &  0.0557 & 3173 / 1672\\
\enddata 
\tablenotetext{a}{BCG included in calibration sample.} 
\end{deluxetable}

\clearpage
\begin{deluxetable}{lllccc}
\tablecolumns{6}
\tablewidth{0pt}
\tablecaption{Control Targets 7\arcsec Radius Photometry\label{table4}}
\tablehead{
\multicolumn{1}{l}{Cluster}
&\multicolumn{2}{c}{Position}
  & \multicolumn{3}{c}{{mAB}}\\
 \multicolumn{1}{l}{}
&\multicolumn{1}{c}{RA}
& \multicolumn{1}{c}{Dec}
  & \multicolumn{1}{c}{\small{FUV}}
  & \multicolumn{1}{c}{\small{NUV}}  
  & \multicolumn{1}{c}{\small{J}\tablenotemark{a}}
  }
\startdata
\multicolumn{6}{c}{\underline{Cluster Ellipticals}} \\  
Abell 389& 02 51 27.1 & -24 57 17.0 &${23.8}^{+0.1}_{-0.1}$ & ${22.90}^{+0.02}_{-0.02}$ & ${16.1}^{+0.02}_{-0.01}$ \\
\nodata & 02 51 22.2  & -24 55 17.0 &${23.4}^{+0.1}_{-0.09}$ & ${22.34}^{+0.02}_{-0.02}$ & ${15.6}^{+0.1}_{-0.1}$ \\
\nodata  & 02 51 23.2  & -24 57 44.0 &${23.6}^{+0.1}_{-0.1}$ & ${22.59}^{+0.02}_{-0.02}$ & ${15.61}^{+0.1}_{-0.09}$ \\
\nodata  & 02 51 33.0  & -24 59 09.3 &${23.06}^{+0.07}_{-0.07}$ & ${22.27}^{+0.02}_{-0.02}$ & ${15.11}^{+0.06}_{-0.06}$ \\
Abell 795  &  09 24 15.4 & +14 07 42.0 &${22.8}^{+0.2}_{-0.2}$ & ${22.59}^{+0.08}_{-0.08}$ & ${15.26}^{+0.06}_{-0.06}$ \\
Abell 2100& 15 36 22.0  & +37 37 43.0 &${24.5}^{+2}_{-0.7}$ & ${22.70}^{+0.07}_{-0.07}$ & ${16.2}^{+0.2}_{-0.1}$ \\
\nodata & 15 36 22.1  & +37 39 34.0 &${24.1}^{+0.9}_{-0.5}$ & ${22.92}^{+0.08}_{-0.08}$ & ${15.9}^{+0.1}_{-0.1}$ \\
Abell 2670  & 23 54 09.1 & -10 24 20.0 & ${23.24}^{+0.1}_{-0.09}$ & ${22.49}^{+0.03}_{-0.02}$ & ${15.37}^{+0.08}_{-0.07}$ \\
\nodata & 23 54 13.2 & -10 23 11.0& ${22.84}^{+0.08}_{-0.07}$ & ${21.86}^{+0.03}_{-0.03}$ & ${14.78}^{+0.04}_{-0.04}$ \\
\nodata & 23 54 18.1 & -10 24 56.0& ${23.3}^{+0.1}_{-0.1}$ & ${22.62}^{+0.06}_{-0.05}$ & ${15.59}^{+0.1}_{-0.09}$ \\
\nodata & 23 54 18.4 & -10 25 09.0& ${22.71}^{+0.07}_{-0.07}$ & ${21.95}^{+0.03}_{-0.03}$ & ${15.10}^{+0.06}_{-0.06}$ \\
\nodata & 23 54 18.9 & -10 23 19.0& ${23.8}^{+0.2}_{-0.1}$ & ${22.86}^{+0.07}_{-0.07}$ & ${16.3}^{+0.2}_{-0.2}$ \\
\nodata & 23 54 15.7 & -10 26 30.0& ${23.06}^{+0.09}_{-0.08}$ & ${21.96}^{+0.03}_{-0.03}$ & ${15.26}^{+0.07}_{-0.07}$ \\
\nodata & 23 54 21.5 & -10 25 11.0& ${21.98}^{+0.04}_{-0.04}$ & ${21.66}^{+0.03}_{-0.03}$ & ${14.71}^{+0.04}_{-0.04}$ \\
\nodata & 23 54 14.3 & -10 21 28.0&${23.20}^{+0.1}_{-0.09}$ & ${22.31}^{+0.04}_{-0.04}$ & ${15.05}^{+0.06}_{-0.05}$ \\
\nodata & 23 54 00.3 & -10 21 45.0& ${22.63}^{+0.07}_{-0.06}$ & ${22.11}^{+0.04}_{-0.04}$ & ${15.32}^{+0.07}_{-0.07}$ \\
\nodata & 23 54 03.5 & -10 20 57.0& ${23.17}^{+0.1}_{-0.09}$ & ${22.52}^{+0.05}_{-0.05}$ & ${15.61}^{+0.1}_{-0.09}$ \\
\multicolumn{6}{c}{\underline{BCGs}} \\
Abell 119  &  00 56 16.1 & -01 15 19.3 &${20.67}^{+0.06}_{-0.06}$ & ${20.23}^{+0.03}_{-0.03}$ & ${13.31}^{+0.01}_{-0.01}$\\
Abell 389& 02 51 24.8  & -24 56 39.0 &${22.52}^{+0.05}_{-0.05}$ & ${21.83}^{+0.02}_{-0.02}$ & ${14.63}^{+0.04}_{-0.04}$\\
Abell 951  & 10 13 50.8 & +34 42 51.1&${23.08}^{+0.08}_{-0.08}$ & ${22.43}^{+0.03}_{-0.03}$ & ${15.24}^{+0.07}_{-0.07}$\\
Abell 963 & 10 17 03.6 & +39 02 49.0 & ${22.32}^{+0.04}_{-0.04}$ & ${22.25}^{+0.03}_{-0.03}$ & ${15.23}^{+0.07}_{-0.06}$\\
Abell 1278  &  11 30 09.1 & +20 30 54.2&${22.7}^{+0.2}_{-0.2}$ & ${22.0}^{+0.1}_{-0.1}$ & ${14.99}^{+0.05}_{-0.05}$\\
Abell 1367  & 11 44 02.2 & +19 56 59.3 & ${19.26}^{+0.03}_{-0.03}$ & ${18.95}^{+0.01}_{-0.01}$ & ${12.285}^{+0.005}_{-0.005}$\\
Abell 1882  &  14 15 08.4 & -00 29 35.7&${22.5}^{+0.3}_{-0.2}$ & ${21.4}^{+0.08}_{-0.08}$ & ${14.58}^{+0.04}_{-0.04}$\\
Abell 2218  & 16 35 49.1 & +66 12 33.9 & ${23.3}^{+0.2}_{-0.2}$ & ${22.47}^{+0.06}_{-0.05}$ & ${15.8}^{+0.1}_{-0.1}$\\
Abell 2219  & 16 40 19.8 & +46 42 41.3& ${23.1}^{+0.3}_{-0.2}$ & ${23.0}^{+0.2}_{-0.2}$ & ${15.8}^{+0.1}_{-0.1}$\\
Abell 2235  & 16 54 43.3 & +40 02 46.4&${22.98}^{+0.07}_{-0.07}$ & ${22.27}^{+0.03}_{-0.02}$ & ${15.35}^{+0.07}_{-0.06}$ \\
Abell 2240  & 16 53 43.7 & +66 45 20.9&${22.7}^{+0.3}_{-0.2}$ & ${22.8}^{+0.3}_{-0.2}$ & ${15.33}^{+0.07}_{-0.06}$\\
Abell 2249 & 17 09 43.8 & +34 24 25.5& ${22.5}^{+0.2}_{-0.2}$ & ${21.83}^{+0.08}_{-0.08}$ & ${14.57}^{+0.03}_{-0.03}$\\
Abell 2255  & 17 12 41.0 & +64 04 21.7& ${23.3}^{+0.6}_{-0.4}$ & ${21.9}^{+0.1}_{-0.1}$ & ${14.95}^{+0.04}_{-0.04}$\\
Abell 2399  & 21 57 29.4 & -07 47 44.6 & ${21.55}^{+0.1}_{-0.09}$ & ${20.78}^{+0.04}_{-0.04}$ & ${13.74}^{+0.02}_{-0.02}$\\
Abell 2448  & 22 31 43.2 & -08 24 31.7 &${21.7}^{+0.2}_{-0.1}$ & ${20.86}^{+0.03}_{-0.03}$ & ${13.66}^{+0.02}_{-0.02}$ \\
Abell 2631  & 23 37 39.7 & +00 16 16.9& ${23.9}^{+0.5}_{-0.4}$ & ${23.6}^{+0.4}_{-0.3}$ & ${16.2}^{+0.2}_{-0.1}$ \\
Abell 2670 & 23 54 13.7 & -10 25 08.0& ${21.18}^{+0.03}_{-0.03}$ & ${20.80}^{+0.02}_{-0.02}$ & ${13.51}^{+0.02}_{-0.02}$ \\
Abell 3158  & 03 42 53.0 & -53 37 52.6 & ${20.77}^{+0.1}_{-0.09}$ & ${20.43}^{+0.05}_{-0.05}$ & ${13.76}^{+0.02}_{-0.02}$ \\
Abell 3330 & 05 14 39.5 & -49 03 29.0 &  ${21.28}^{+0.03}_{-0.03}$ & ${20.517}^{+0.007}_{-0.007}$ & ${13.85}^{+0.02}_{-0.02}$ \\
Abell 3376  &  06 02 09.71 & -39 56 59.5 & ${20.77}^{+0.09}_{-0.08}$ & ${20.43}^{+0.04}_{-0.04}$ & ${13.83}^{+0.02}_{-0.02}$\\
Abell 3556  & 13 24 06.7 & -31 40 12.1&${21.1}^{+0.1}_{-0.1}$ & ${20.01}^{+0.04}_{-0.04}$ & ${13.04}^{+0.01}_{-0.01}$\\
Abell 3716  & 20 51 56.9 & -52 37 47.2 &${21.2}^{+0.1}_{-0.1}$ & ${20.39}^{+0.03}_{-0.03}$ & ${13.44}^{+0.02}_{-0.02}$ \\
\enddata 
\tablenotetext{a}{From the 2MASS Database.}
\end{deluxetable}

\clearpage
\begin{deluxetable}{ccccccc}
\tablecolumns{7}
\tablewidth{0pc}
\tablecaption{Control Sample Fitting Parameters\label{table5}}
\tablehead{                
\colhead{}          &
\multicolumn{3}{c}{FUV} & 
\multicolumn{3}{c}{NUV} \\
\colhead{Fit} &
 \colhead{$C_1$} & 
 \colhead{$C_2$} &
 \colhead{Scatter} &
\colhead{$C_1$} &
 \colhead{$C_2$} &
 \colhead{Scatter}
}
\startdata
 {${\rm{L_{UV}}-{\rm{L_{J}}}}$}& $ -1.16\pm{0.023}$ & $ 1.00\pm{ 0.062}$&  0.10 & $ -0.85\pm{0.016}$ & $ 0.89\pm{ 0.045}$&  0.08  \\
  {$f_{\nu{\rm{,UV}}}/f_{\nu{\rm{,J}}}-{\rm{L_{J}}}$}&  $-3.05\pm{ 0.028}$ & $ -0.08\pm{ 0.075}$&  0.13 & $-2.77\pm{ 0.016}$ & $ -0.14\pm{ 0.046}$&  0.08 \\
\enddata
\tablecomments{Fits are described as Y-X (see Equation~\ref{eq:powerlaw}). UV luminosity is in units of $10^{43}$ erg s$^{-1}$ and J band luminosity is in units of $10^{44}$ erg s$^{-1}$.}
\end{deluxetable}

\clearpage
\begin{deluxetable}{lcccccc}
\rotate
\tablecolumns{5}
\tablewidth{0pt}
\tablecaption{UV Excess and Estimated Star Formation Rates\label{table6}}
\tablehead{
\multicolumn{1}{l}{Cluster}
&\multicolumn{1}{c}{${\rm{L}}_{\rm{FUV}}$ Excess}
& \multicolumn{1}{c}{Cont. SFR\tablenotemark{a}}
& \multicolumn{1}{c}{Burst Stellar Mass\tablenotemark{b}}
  & \multicolumn{1}{c}{{${\rm{L}}_{\rm{NUV}}$ Excess}}
  & \multicolumn{1}{c}{{Cont. SFR\tablenotemark{a}}}
& \multicolumn{1}{c}{Burst Stellar Mass\tablenotemark{b}}  \\
 \multicolumn{1}{l}{}
&\multicolumn{1}{c}{[$10^{43}$ erg ${\rm{sec}}^{-1}$]}
& \multicolumn{1}{c}{[\msun~${\rm{yr}}^{-1}$]}
& \multicolumn{1}{c}{[$10^6$ \msun]}
  & \multicolumn{1}{c}{{[$10^{43}$ erg ${\rm{sec}}^{-1}$]}}
  & \multicolumn{1}{c}{{[\msun~${\rm{yr}}^{-1}$]}}  
& \multicolumn{1}{c}{[$10^6$ \msun]}
  }
\startdata
Abell 85 &   ${0.026}\pm{0.006}$  &${0.031}$ & $1.2$&${0.037}\pm{0.008}$ & ${0.05}$&$1.7$\\
Abell 644 &   \nodata  &\nodata & \nodata&${0.003}\pm{0.01}$ & ${0.004}$&$0.13$\\
Abell 1204 &   ${0.39}\pm{0.04}$  &${0.41}$ & $15.9$&${0.31}\pm{0.03}$ & ${0.34}$&$12.7$\\
Abell 2029 &   ${0.03}\pm{0.01}$  &${0.03}$ & $1.3$&${0.05}\pm{0.02}$ & ${0.06}$&$2.1$\\
Abell 2052 &   ${0.009}\pm{0.003}$  &${0.011}$ & $0.4$&${0.013}\pm{0.005}$ & ${0.017}$&$0.6$\\
Abell 2142 & ${0.008}\pm{0.01}$ & ${0.009}$ &  $0.3$& $<0.01$  &$<0.01$&$<0.5$\\
Abell 2597 &   ${0.69}\pm{0.04}$  &${0.81}$ & $31.2$&${0.88}\pm{0.05}$ & ${1.07}$&$38.7$\\
Abell 3112 &   ${0.08}\pm{0.01}$  &${0.09}$ & $3.5$&${0.09}\pm{0.01}$ & ${0.10}$&$3.8$\\
Hercules A &   ${0.20}\pm{0.04}$  &${0.22}$ &$8.4$ &${0.27}\pm{0.04}$ & ${0.30}$&$11.1$\\
Hydra A   &   ${0.68}\pm{0.04}$  &${0.82}$ & $31.2$&${0.97}\pm{0.05}$ & ${1.22}$&$44.0$\\
MKW3s   & ${0.009}\pm{0.004}$ & ${0.011}$& $0.4$  &$<0.005$  &$<0.007$& $<0.2$\\
MKW4   &   ${0.013}\pm{0.002}$  &${0.016}$ & $0.6$&${0.007}\pm{0.004}$ & ${0.009}$&$0.3$\\
MS0839.8+2938   &   ${1.03}\pm{0.08}$  &${1.05}$& $40.7$& ${1.20}\pm{0.09}$ & ${1.25}$&$46.7$\\
MS1358.4+6245  &   ${0.21}\pm{0.07}$  &${0.19}$& $7.8$& ${0.14}\pm{0.06}$ & ${0.13}$&$4.8$\\
MS1455.0+2232  &   ${3.2}\pm{0.2}$  &${3.0}$& $119$& ${3.0}\pm{0.2}$ & ${2.9}$&$110$\\
RXJ1347.5-1145  &   ${4.0}\pm{0.3}$  &${3.5}$& $157$& ${5.9}\pm{0.4}$\tablenotemark{c} & ${4.6}$\tablenotemark{c}&$177$\\
ZwCl 3146  &   ${11.3}\pm{0.6}$\tablenotemark{c}  &${10.4}$\tablenotemark{c} & $413$&${9.4}\pm{0.5}$ & ${8.9}$&$337$\\
\enddata 
\tablenotetext{a}{Assuming continuous star formation over 20 Myr.}
\tablenotetext{b}{Lower limit of total star formation obtainable from the single burst assumption, set by a starburst occurring 10 Myr ago.}
\tablenotetext{c}{Some Ly$\alpha$ contamination is suspected.}
\end{deluxetable}

\clearpage
\begin{deluxetable}{lcccccccc}
\rotate
\tablecolumns{9}
\tablewidth{0pt}
\tablecaption{Multiwavelength Comparison Data\label{table7}}
\tablehead{
\multicolumn{1}{l}{Cluster}
&\multicolumn{1}{c}{${\rm{L}}_{{\rm{H}}\alpha}$ (Ref.)}
  & \multicolumn{1}{c}{{${\rm{L}}_{\rm{IR}}$} (Ref.)}
& \multicolumn{1}{c}{IR SFR\tablenotemark{a}}
&\multicolumn{1}{c}{IRX\tablenotemark{b}}
  & \multicolumn{2}{c}{Entropy\tablenotemark{c} [keV ${\rm{cm}}^{2}$]}
  & \multicolumn{2}{c}{Cooling Time\tablenotemark{c} [Gyr]}\\
  \multicolumn{1}{l}{}
&\multicolumn{1}{c}{[$10^{40}$ erg ${\rm{sec}}^{-1}$]}
&\multicolumn{1}{c}{[$10^{44}$ erg ${\rm{sec}}^{-1}$]}
& \multicolumn{1}{c}{[\msun~${\rm{yr}}^{-1}$]}
&\multicolumn{1}{c}{}
  & \multicolumn{1}{c}{Central}
  & \multicolumn{1}{c}{R$=20$ kpc}
  & \multicolumn{1}{c}{Central}
  & \multicolumn{1}{c}{R$=20$ kpc}
  }
\startdata
Abell 85 &   $<{0.43}$ (1)  &${0.28}\pm{0.03}$ (7) & ${1.6}\pm{0.2}$ & ${1.1}\pm{0.1}$ &$12.5\pm{0.5}$&${38}\pm{1}$& $0.313\pm{0.009}$&${1.18}\pm{0.04}$\\
Abell 644 &   \nodata &\nodata &\nodata& \nodata &${132}\pm{9}$&${153}\pm{21}$& ${4}\pm{2}$&${4.6}\pm{0.8}$\\
Abell 1204 &   ${6.5}\pm{0.8}$ (2)  &${1.7}\pm{0.2}$ (7) & ${8.1}\pm{0.8}$ &  ${1.1}\pm{0.1}$ &${15}\pm{1}$&${23}\pm{1}$& ${0.38}\pm{0.02}$&${0.59}\pm{0.04}$\\
Abell 2029 &   $<{0.64}$ (3) &\nodata & \nodata & \nodata &${10.5}\pm{0.7}$&${50}\pm{2}$&${0.238}\pm{0.006}$& ${1.03}\pm{0.04}$\\
Abell 2052 &   ${1.5}\pm{0.1}$ (2)  &${0.24}\pm{0.02}$ (7) & ${1.4}\pm{0.1}$ &  ${1.4}\pm{0.2}$ &${9.4}\pm{0.7}$&${34}\pm{1}$& ${0.53}\pm{0.04}$&${1.07}\pm{0.04}$\\
Abell 2142 &   $<{0.90}$ (1)  &\nodata & \nodata & \nodata &${68}\pm{2}$&${84}\pm{4}$&${2.0}\pm{0.1}$& ${2.5}\pm{0.1}$\\
Abell 2597 &   ${51}\pm{7}$ (1)  &${0.93}\pm{0.09}$ (8) & ${3.4}\pm{0.3}$ &  ${0.60}\pm{0.07}$ &${11}\pm{2}$&${26}\pm{2}$& ${0.31}\pm{0.02}$&${0.69}\pm{0.06}$\\
Abell 3112 &   ${3.8}\pm{0.5}$ (1) &${0.84}\pm{0.08}$ (7) & ${4.2}\pm{0.4}$ & ${1.3}\pm{0.2}$ &${11}\pm{1}$&${37}\pm{3}$& ${0.33}\pm{0.02}$&${1.01}\pm{0.08}$\\
Hercules A &   ${1.3}\pm{0.2}$ (1)  &$<{2.2}$ (9) & $<{7.56}$ &  $<{1.4}$ &${9}\pm{1}$&${33}\pm{1}$& ${0.48}\pm{0.02}$&${1.31}\pm{0.06}$\\
Hydra A   &   ${10}\pm{1}$ (1) &${0.5}\pm{0.2}$ (10) & ${2.1}\pm{0.2}$ &  ${0.4}\pm{0.1}$ &${13.3}\pm{0.7}$&${34}\pm{1}$& ${0.43}\pm{0.02}$&${1.10}\pm{0.04}$\\
MKW3s   &   ${0.50}\pm{0.07}$\tablenotemark{d} (4)  &\nodata & \nodata & \nodata &${24}\pm{2}$&${51}\pm{2}$& ${0.82}\pm{0.09}$&${1.8}\pm{0.1}$\\
MKW4   &   \nodata  &$<{0.033}$ (11) & $<{0.21}$ &  $<{0.5}$ &${6.9}\pm{0.3}$&${58}\pm{4}$& ${0.101}\pm{0.003}$&${2.2}\pm{0.2}$\\
MS0839.8+2938   &   ${67}\pm{9}$ (5)  &\nodata & \nodata & \nodata &${19}\pm{3}$&${32}\pm{3}$& ${0.57}\pm{0.06}$&${1.0}\pm{0.1}$\\
MS1358.4+6245  &   ${4.9}\pm{0.4}$ (6)  &${1.4}\pm{0.1}$ (12) & ${4.87}\pm{0.5}$ &  ${1.1}\pm{0.2}$ &${21}\pm{3}$&${31}\pm{3}$& ${0.68}\pm{0.07}$&${1.0}\pm{0.1}$\\
MS1455.0+2232  &   ${33}^{+4}_{-2}$ (2)  &${2.0}\pm{0.2}$ (12) & ${6.81}\pm{0.7}$ &  ${0.26}\pm{0.03}$ &${17}\pm{2}$&${26}\pm{2}$&${0.40}\pm{0.02}$& ${0.59}\pm{0.04}$\\
RXJ1347.5-1145  & \nodata &$<{84}$ (10) & $<{350}$ &  $<{1.8}$ &${12}\pm{21}$&${46}\pm{8}$& ${0.44}\pm{0.06}$&${0.62}\pm{0.09}$\\
ZwCl 3146  &   ${458}^{+11}_{-8}$ (2) &${15} \pm{2}$ (12)& ${65}\pm{6}$ & ${0.62}\pm{0.09}$ &${11}\pm{2}$&${24}\pm{2}$& ${0.36}\pm{0.02}$&${0.52}\pm{0.04}$\\
\enddata 
\tablecomments{Spitzer IR flux uncertainties are conservatively estimated at 10\%.}
\tablenotetext{a}{Calculated from total IR luminosities following the method of~\citet{odea08}.}
\tablenotetext{b}{IR excess determined as in~\citet{johnson06}.}
\tablenotetext{c}{Values from~\citet{cavagnolo09}}
\tablenotetext{d}{15\% errors are assumed.}
\tablerefs{(1)~\citet{cavagnolo09}; (2)~\citet{crawford99}; (3)~\citet{jaffe05}; (4)~\citet{salome03}; (5)~\citet{donahue92}; (6)~\citet{lamareille06}; (7)~\citet{odea08}; (8)~\citet{donahue07}; (9)~\citet{golombek88}; (10)~\citet{edge01}; (11)~\citet{knapp89}; (12)~\citet{egami06}}
\end{deluxetable}


\end{document}